\documentclass[prd,amsmath,amssymb,twocolumn,superscriptaddress]{revtex4-2}
\usepackage{graphicx}	
\usepackage{subfigure}	

\newcommand\nn{\nonumber}
\newcommand\eq[1] {\begin{align} #1 \end{align}}

\newcommand{\br}[1]{\left( #1 \right)}
\newcommand{\brs}[1]{\left[ #1 \right]}

\newcommand{\brm}[1]{\left| #1 \right|}

\newcommand{\Sp}{\mbox{Tr}}

\newcommand{\vv}[1]{{\bf #1}}

\newcommand{\dd}[1]{{\hat #1}}        

\newcommand{\M} {{\cal M}} 

\newcommand{\GeV}{\mbox{GeV}}
\newcommand{\MeV}{\mbox{MeV}}

\newcommand{\rad}{\mbox{rad}}

\allowdisplaybreaks		

\begin{document}

\title{Exact calculation of photon polarization observables in Bethe--Heitler process}


\author{Yu.\,M. Bystritskiy}
\affiliation{Joint Institute for Nuclear Research \\
Bogoliubov Laboratory of Theoretical Physics, 141980 Dubna, Russia}

\author{V.\,A. Zykunov}
\affiliation{Joint Institute for Nuclear Research \\
Bogoliubov Laboratory of Theoretical Physics, 141980 Dubna, Russia}
\affiliation{Francisk Skorina Gomel State University, 246028 Gomel, Belarus}

\date{\today}

\begin{abstract}
The effects of polarization transfer from the initial electron to the bremsstrahlung photon in
electron–nucleus scattering (Bethe--Heitler process) are considered.
The calculation is carried out without the assumption of smallness of the electron mass.
A detailed comparison with previous well-known works is made.
Some issues related to neglecting the electron mass are shown and commented on.
The results are applicable to the modelling of polarized cross sections at low energies
(even at a few electron mass scale).
\end{abstract}

\maketitle

\section{Introduction}

As a basic process of quantum electrodynamics (QED), the bremsstrahlung reaction has been investigated
since the early days of QED~\cite{Sommerfeld:1931qaf,Heitler:1933giw,Bethe:1934za}. It was only in the fifties
that polarization phenomena in this process were calculated \cite{Wick:1951,May:1951a,May:1951b}.
More careful calculations of the polarization of bremsstrahlung radiation induced by
an electron beam in the vicinity of a nuclear field \cite{Olsen:1957ahf,Olsen:1959zz} led to the
idea of producing polarized photon beams.
It was shown that an unpolarized electron beam generates linearly polarized photons,
while a polarized electron beam generates circularly polarized photons with polarization proportional
to the initial electron beam polarization.
These features were extensively used at numerous accelerator facilities, for instance,
in the experimental Hall B at the Thomas Jefferson National Accelerator Facility (JLab)
to operate a high energy polarized photon beam \cite{CLAS:2003umf} and more recently in the experimental
Hall D of JLab hosting the GlueX experiment \cite{GlueX:2020idb} designed to search for hybrid mesons
with a $9~\GeV$ linearly polarized photon beam.

Scientific interest in polarization of photons is not limited to its filtering power
with respect to some reaction mechanisms, but also extends to the production of secondary
beams, in particular polarized positron beams. It is indeed predicted that a circularly
polarized photon beam of sufficient energy can create a polarized $e^+e^-$ pair whose
longitudinal and transverse polarization components are both proportional to the initial photon beam
polarization~\cite{Olsen:1959zz}. Polarization transfer in the transverse plane is
however much less efficient than in the longitudinal plane, resulting in an essentially
longitudinally polarized positron beam. As compared to earlier calculations,
the experimental demonstration of the circular--to--longitudinal polarization transfer
is relatively recent and was carried out at
the High Energy Accelerator Research Organization (KEK)~\cite{Omori:2005eb},
the Stanford Linear Accelerator Center (SLAC)~\cite{Alexander:2008zza}, and
the Continuous Electron Beam Accelerator Facility (CEBAF)~\cite{PEPPo:2016saj}
using completely different techniques to produce the initial polarized photon beam.
Particularly, the proof--of--principle PEPPo experiment~\cite{Voutier:2017ihu} demonstrated
an efficient transfer of the polarization of the initial electron beam to
a secondary positron beam generated by polarized bremsstrahlung
of the initial electrons \cite{Potylitsyn:1997wn}. This technique, particularly suited to continuous electron
facilities, will extend polarized positron capabilities from MeV to GeV electron
accelerators and open access to physics studies with polarized positron beams to a wide community.

The perspective of operating high intensity polarized positron beam at CEBAF
generated intense scientific activity, which is summarized in a recent
Topical Issue of the European Physics Journal~\cite{Alamanos:2022wwn}.
It develops a wide experimental program~\cite{Accardi:2020swt} based on a positron source
expected to deliver a continuous positron beam, either polarized or unpolarized
with polarization as high as 60\% and intensity ranging from 100~nA up to 3~$\mu$A
inversely dependent on the beam polarization degree.
This would be achieved with a new positron injector \cite{Habet:2022fch} based on the PEPPo technique.

The optimization of the design of such a facility is an involved process which starts
with an accurate and unrestricted description of a polarization transfer of bremsstrahlung
and pair creation reactions. Earlier calculations \cite{Olsen:1959zz}
developed in the  ultra-relativistic and small angle approximations were shown of
restricted applications \cite{Dumas:2009vy} since the electron mass cannot be neglected
over the full kinematic domain: specifically, close to the bremsstrahlung end--point
when the electron transfers all its kinetic energy to the photon,
and conversely, when one of the pair particles is produced close to rest.
These limitations are particularly important when the energy of the initial
electron (photon) is small \cite{Dumas:2011can}. These features were corrected
in the finite lepton mass calculations in Ref.~\cite{Kuraev:2009uy},
which were worked--out in the peripheral kinematics limit.
There are other interesting kinematical domains where the influence of the finite electron mass
is important. For example, in Ref.~\cite{Barbaro:2013waa} some of these domains were shown
to be crucial for restoring the mass dependence of unpolarized observables.

The purpose of the present work is to derive and discuss complete expressions for
polarization observables of the bremsstrahlung reaction that are free of any kinematical
restrictions and derived at the first order of perturbation theory.
These calculations are aimed at optimizing the polarized photon
production, for instance, for the purposes of a PEPPo-like polarized positron
source at low energies.

The paper is organized in the following way.
In Section~\ref{sec.Kinematics} the main kinematic parameters of the scattering process are defined.
Section~\ref{sec.CrossSection} gives the amplitude of the process with
some discussion of the initial and final particle spin states. The cross sections of
unpolarized and polarized particle collisions are also defined in this section.
Section~\ref{sec.BeyondBornApproximation} is devoted to the discussion
of different aspects of the process, which appear beyond the Born approximation.
In Section~\ref{sec.Comparison} we present the comparison of our precise calculation
with previous results.
Section~\ref{sec.NumericalAnalysis} shows different spectral distributions
and the behaviour of the cross sections as a function of different kinematic
variables at different energies.
Section~\ref{sec.Issues} is dedicated to the description of a set of
physical issues that manifests themselves when using known formulae beyond the scope of
their applicability. Two appendices contain useful expressions.
In Appendix~\ref{sec.PhotonPolarizationVectors} we construct pure polarization photon states
and in Appendix~\ref{sec.ExplicitCrossSections} the complete exact expressions for unpolarized
and polarized cross sections are presented.

\section{Kinematics}
\label{sec.Kinematics}

\begin{figure}
\centering
\includegraphics[width=0.2\textwidth]{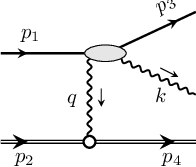}
\caption{The notation of momenta for the bremsstrahlung process.}
\label{fig.Process}
\end{figure}

Let us consider the process of bremsstrahlung of one single photon during the collision of an electron with a nucleus,
see Fig.~\ref{fig.Process}. The 4-momentum conservation law then reads as
\eq{
 p_1 + p_2 = p_3 + p_4 + k,
\label{eq.MomentumConservation}
}
where $p_{1,3} = \br{E_{1,3}, \vv{p_{1,3}}}$ are the 4-momenta
of the initial and scattered electrons with energies $E_{1,3}$ and 3-momenta $\vv{p_{1,3}}$;
$p_{2,4} = \br{E_{2,4}, \vv{p_{2,4}}}$ are the 4-momenta of the initial and recoil nucleus
with energies $E_{2,4}$ and 3-momenta $\vv{p_{2,4}}$; and $k = \br{\omega, \vv{k}}$ is the bremsstrahlung photon 4-momentum,
where $\omega$ stands for its energy and $\vv{k}$ is for its 3-momentum.
We develop our formalism in the laboratory system, considering
the target to be at rest {\it i.e.} $\vv{p_2} = \vv{0}$ and $E_2 = M_A$, where $M_A$ is the mass of the target nucleus.
Thus, the total energy of the reaction is defined by the energy of the initial electron $E_1$ with mass $m_e$.
The square of the total center-of-mass energy $s$ is
\eq{
	s = \br{p_1 + p_2}^2 = M_A \br{ 2 E_1 + M_A } + m_e^2.
}
The 4-momentum transferred from the incoming electron to the target is
$q = p_4 - p_2$ and its square is
\eq{
	q^2  = 2 M_A \br{M_A - E_4} = - \frac{2 M_A}{M_A + E_4}\brm{\vv{p}_4}^2.
}
Let us note that the second form of this expression is more suitable for numerical
treatment since it does not contain the difference of two large but almost equal quantities
(at low momentum transfer the final nucleus moves slowly and thus $E_4 \approx M_A$).

\begin{figure}
\centering
\includegraphics[width=0.4\textwidth]{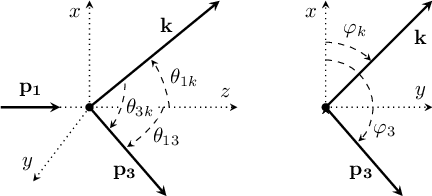}
\caption{The disposition of the momenta in the laboratory frame system.
Side view (left) and view along the beam (right).}
\label{fig.Angles}
\end{figure}

The phase volume of the reaction in Fig.~\ref{fig.Process} has the
standard form
\eq{
	\!d^9\Phi_3
	&=
	\frac{1}{\br{2\pi}^5} \delta\br{p_1 + p_2 - p_3 - p_4 - k}
	\frac{d\vv{p_3}}{2 E_3}
	\frac{d\vv{p_4}}{2 E_4}
	\frac{d\vv{k}}{2 \omega}.
	\label{eq.PhaseVolume0}
}
Taking into account the $\delta$-function and integrating over trivial variables, we get:
\eq{
	d\Phi_3
	&=
	\frac{1}{2^8 \pi^5} F_b~
	\omega \, d\omega \, d\Omega_{\vv{k}} d\Omega_{\vv{p_3}},
	\label{eq.PhaseVolume}
}
where $d\Omega_{\vv{k}} = dC_{1k} \, d\varphi_k$
and $d\Omega_{\vv{p_3}} = dC_{13} \, d\varphi_3$ are the solid angles of the final photon and the scattered electron
(we use the following notation for the cosines of the angles
$C_{1k} \equiv \cos \theta_{1k}$ and $C_{13} \equiv \cos \theta_{13}$).
The reaction angles $\theta_{1k}$ and $\theta_{13}$ are the angles between the initial electron beam momentum $\vv{p_1}$ and the momenta of the bremsstrahlung photon $\vv{k}$ and the scattered electron $\vv{p_3}$, respectively (see Fig.~\ref{fig.Angles}). All the azimuthal angles are measured from the $x$-axis, see the right panel
of Fig.~\ref{fig.Angles} where we present the configuration of momenta from the beam point of view.
It is important to note here that in the literature it is common to select some plane
to measure all azimuthal angles with respect to it.
Say, if one is interested in the final electron spectra or angular distribution, then
it is convenient to define the {\it scattering plane} as the plane defined by the initial electron momentum $\vv{p_1}$ and
by the scattered electron momentum $\vv{p_3}$.
In Ref.~\cite{Olsen:1959zz}, the momenta $\vv{p_1}$ and $\vv{k}$ lie in the
{\it radiation plane} or the {\it plane of emission}.
If the initial electron beam has no transversal polarization, then there is no selected azimuthal direction,
and the dependence on the overall azimuthal orientation of the process is trivial. For example,
the integration over the final photon azimuthal angle then gives just
\eq{
	\int\limits_0^{2\pi} d\varphi_k \cdots \rightarrow 2\pi \cdots.
}
We discuss this aspect in more detail in Section~\ref{sec.NumericalAnalysis}.
Here we keep this freedom for convenience: the orientation of the $x$-axis is arbitrary and selected for the general convenience, and all the azimuthal angles are then measured from this $x$-axis.

The dimensionless quantity $F_b$ in (\ref{eq.PhaseVolume}) is the Jacobian of mapping to the new variables,
and it has the form
\eq{
	F_b = \frac{\brm{\vv{p_3}}^2}
	{\brm{\vv{p_3}} E_4 - E_3 \brm{\vv{p_4}} C_{34}},
	\label{eq.Fb}
}
where the cosine $C_{34}$ between the momenta $\vv{p_3}$ and $\vv{p_4}$ reads as:
\eq{
 	C_{34}  &= \frac{ \brm{\vv{p_1}} C_{13} - \brm{\vv{p_3}} - \brm{\vv{k}} C_{3k} }{\brm{\vv{p_4}}}.
}
For completeness, we also present the cosine $C_{14}$ between the momenta $\vv{p_1}$ and $\vv{p_4}$:
\eq{
 	C_{14}  &= \frac{ \brm{\vv{p_1}} - \brm{\vv{p_3}} C_{13} - \brm{\vv{k}} C_{1k} }{\brm{\vv{p_4}}}.
}
The energy of the scattered electron $E_3$ is fixed by the energy--momentum conservation law (\ref{eq.MomentumConservation})
(or mathematically by the $\delta$-function in (\ref{eq.PhaseVolume0})) and is equal to:
\eq{
	E_3 &= \frac{-B + c_2 \sqrt{D}}{A},
	\label{eq.E3}
}
where $A = c_1^2 - c_2^2$, $B = c_1 c_3$,
$D = c_3^2 - A m_e^2$
and
\eq{
	c_1 &= E_1 - \omega + M_A,
	\nn\\
	c_2 &= \brm{\vv{p_1}} C_{13} - \brm{\vv{k}} C_{3k},
	\\
	c_3 &= E_1 \br{\omega - M_A} + \omega M_A - \brm{\vv{p_1}} \brm{\vv{k}} C_{1k} - m_e^2.
	\nn
}
The expression for $C_{3k}$ (the cosine of the angle $\theta_{3k}$ between the
momenta $\vv{p_3}$ and $\vv{k}$)
can be found using the cosine law as
\eq{C_{3k} = C_{13} C_{1k} + S_{13} S_{1k} \cos \br{\varphi_3 - \varphi_k},
}
where $S_{13} \equiv \sin \theta_{13}$ and $S_{1k} \equiv \sin \theta_{1k}$.

\section{Cross section}
\label{sec.CrossSection}

\begin{figure}
	\centering
    \subfigure[]{\includegraphics[width=0.19\textwidth]{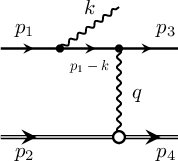}\label{fig.DiagramA}}
	\hspace{0.05\textwidth}
    \subfigure[]{\includegraphics[width=0.19\textwidth]{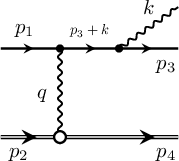}\label{fig.DiagramB}}
    \caption{
    	The Feynman diagrams of the Bethe-Heitler process.
    }
    \label{fig.Diagrams}
\end{figure}

In the lowest order of perturbation theory the Bethe-Heitler process is described by
the one--photon exchange approximation.
Two Feynman diagrams correspond to this approximation, see Fig.~\ref{fig.Diagrams}.
We neglect the diagrams with emission of photons from the nucleus
since the mass of the nucleus $M_A$ is much larger than the mass of the electron $m_e$, and
we are mostly interested in the low-energy regime.

Since our task is to get the cross section for the emission of a polarized photon
(whose polarization is parameterized by the Stokes parameters), we need to
calculate the bremsstrahlung of the photon in each of two specific pure polarization states $\varepsilon_{1,2}^\mu$
and then assemble the Stokes parameters combining these pure polarized cross sections \cite{McMaster:1954}.
The explicit construction of these pure polarization states $\varepsilon_{1,2}^\mu$ is
given in Appendix~\ref{sec.PhotonPolarizationVectors}.
In order to get the invariant amplitudes of the process, we apply the Feynman rules of QED to the diagrams in Fig.~\ref{fig.Diagrams} and get the following expression:
\eq{\M_i =
	\frac{e^3 Z}{q^2}
	[ \bar{u}(p_3) O_{\mu\nu} u(p_1) ]
	[ \bar{u}(p_4) \gamma^\mu u(p_2) ] \,
	\varepsilon_i^\nu.
	\label{eq.Amplitude}
}
where the quantity $Z$ is the electric charge of the nucleus in
the units of the proton charge $e = \sqrt{4\pi \alpha} > 0$.
The value of fine structure constant $\alpha \approx 1/137$ and other
parameter values are further taken from the PDG summary \cite{Zyla:2020zbs}.
The quantity $O_{\mu\nu}$ in (\ref{eq.Amplitude}) has the form
\eq{
	O_{\mu\nu}
	=
	\frac{\gamma_\mu (\dd{p_1}-\dd{k} + m_e) \gamma_\nu}{-2(p_1 k)}
	+
	\frac{\gamma_\nu (\dd{p_3}+\dd{k} + m_e) \gamma_\mu}{2(p_3 k)},
	\label{eq.OperatorOdefinition}
}
where the notation $\dd{p} \equiv \gamma_\mu p^\mu$ is used.
These two terms here correspond to two diagrams in Fig.~\ref{fig.Diagrams}.

The cross section of the process can be written in the standard form:
\eq{
	d\sigma_{ij} = \frac{1}{4 F} \overline{\sum_{s}} \M_i \M_j^+ d\Phi_3,
	\label{eq.CrossSection}
}
where the phase volume of the final particles $d\Phi_3$ was defined
in (\ref{eq.PhaseVolume}) and
the invariant flux $F$ of the initial particles can be written as:
\eq{
	F = \sqrt{\br{p_1 p_2}^2 - m_e^2 M_A^2} = \frac{1}{2}\lambda^{\frac{1}{2}}(s, m_e^2, M_A^2),
	\label{eq.Flux}
}
here $\lambda(x,y,z) = x^2 + y^2 + z^2 - 2 x y - 2 x z - 2 y z$ is the well-known triangle function.

The overline summation operator $\overline{\sum}$ in (\ref{eq.CrossSection})
indicates the average over the initial particle spin states and
the sum over the final particle spin states (except for photon polarizations):
\eq{
	\overline{\sum} \cdots
	=
	\frac{1}{2} \sum_{s_1}
	\brs{
		\frac{1}{2} \sum_{s_2}
		\brs{
			\sum_{s_3, s_4} \cdots
	}}.
	\label{eq.SumAveragingOperator}
}
In order to perform this operation for fermions, we use the spin density matrices in the form:
\eq{
	\frac{1}{2} \sum_{s} u^{s}(p_1) \bar{u}^{s}(p_1) &= \frac{1}{2} \br{\dd{p_1} + m_e} \br{1 - \gamma_5 \dd{a_1}},
	\nn\\
	\frac{1}{2} \sum_{s} u^{s}(p_2) \bar{u}^{s}(p_2) &= \frac{1}{2} \br{\dd{p_2} + M_A},
	\label{eq.SpinSummationOfFermions}\\
	\sum_{s} u^{s}(p_3) \bar{u}^{s}(p_3) &= \dd{p_3} + m_e,
	\nn\\
	\sum_{s} u^{s}(p_4) \bar{u}^{s}(p_4) &= \dd{p_4} + M_A,
	\nn
}
where $a_1 = \br{ a_1^0, a_1^x, a_1^y, a_1^z }$ is the initial electron beam polarization vector.
For {\it a longitudinally} polarized electron beam one has:
\eq{
	a_1 &= \lambda_1 \br{ \frac{\brm{\vv{p_1}}}{m_e}, 0, 0, \frac{E_1}{m_e} },
	\label{eq.LongitudinalElectronPolarization}
}
where $\lambda_1 = \pm 1$ is the doubled projection of the electron spin to the $z$-axis direction
for an electron at rest.
In the case of {\it transversal polarization} of the electron beam, the vector $a_1$ is the unit vector perpendicular
to the beam direction (it lies in the $xy$-plane) and its orientation is fixed by the azimuthal angle $\varphi_a$
which is measured from the $x$-axis:
\eq{
	a_1 &= \br{ 0, \cos\varphi_a, \sin \varphi_a, 0 }.
	\label{eq.TransverseElectronPolarization}
}

The product of the amplitudes in (\ref{eq.CrossSection}) after the action of the sum operator $\overline{\sum}$ then reads as:
\eq{
	\overline{\sum_{s}} \M_i \M_j^+
	&=
	\frac{\br{4\pi\alpha}^3 Z^2}{4 (q^2)^2} S_{ij},
	\label{eq.Msquare}
}
where the quantity $S_{ij}$ is expressed in terms of the traces of the product of Dirac $\gamma$-matrices:
\eq{
	S_{ij}
	&=
	\varepsilon_i^\alpha (\varepsilon_j^\beta)^*
	\times\nn\\
	&\hspace{2mm}\times
	\Sp \brs{ (\dd{p_1} + m_e ) \br{1 - \gamma_5 \dd{a_1}} \widetilde O_{\nu\beta} (\dd{p_3} + m_e ) O_{\mu\alpha} }
	\times\nn\\
	&\hspace{2mm}\times
	\Sp \brs{ (\dd{p_2} + M_A) \gamma^\nu (\dd{p_4} + M_A) \gamma^\mu }.
	\label{eq.SbDefinition}
}
Here the tilde sign denotes the Dirac conjugation, i.e. $\widetilde O \equiv \gamma_0 O^+ \gamma_0$, where $O^+$ is the Hermitian conjugation.
The evaluation of these traces leads to rather cumbersome expressions.
The explicit expressions for the quantities $S_{ij}$ are presented in Appendix~\ref{sec.ExplicitCrossSections}.

Let us now obtain the final expression for the cross section used in numerical estimation.
Substituting the averaged amplitudes from (\ref{eq.Msquare}) and the phase volume $d\Phi_3$
from (\ref{eq.PhaseVolume}) into the general formula (\ref{eq.CrossSection}), one gets:
\eq{
	d\sigma_{ij}
	=
	\frac{\alpha^3 Z^2}{2^6 \pi^2 F}
	\frac{\omega F_b}{(q^2)^2}
	S_{ij}
	d\omega \, d\Omega_{\vv{k}} d\Omega_{\vv{p_3}}.
	\label{eq.FinalCrossSection}
}
It is convenient to parameterize the polarization state of the final photon
in terms of the Stokes parameters $P_{1,2,3}$ \cite{McMaster:1954}.
In order to do that, we assemble the cross sections $d\sigma_{ij}$
into the following combinations:
\eq{
	d\sigma_I 	  &= d\sigma_{11} + d\sigma_{22},
	\label{eq.def.SigmaI}\\
	d\sigma_{P_1} &= d\sigma_{11} - d\sigma_{22},
	\label{eq.def.SigmaP1}\\
	d\sigma_{P_2} &= d\sigma_{12} + d\sigma_{21},
	\label{eq.def.SigmaP2}\\
	d\sigma_{P_3} &= i \br{ d\sigma_{21} - d\sigma_{12} },
	\label{eq.def.SigmaP3}
}
where we follow the notation from \cite{McMaster:1954} (see formula (27) there).
The quantity $d\sigma_I$ from (\ref{eq.def.SigmaI}) is the cross section for
the unpolarized photon production corresponding to the overall intensity $I$ of the
produced photon beam. Then the Stokes parameters $P_i$ of the polarized photon beam
are defined by the following expressions:
\eq{
	d\sigma_{P_i} = P_i ~ d\sigma_I, \hspace{10mm} i=1,2,3.
}

\section{Beyond Born approximation}
\label{sec.BeyondBornApproximation}

The model dependent part of the formulae above are given in the frame of the
Born approximation, which corresponds to the one-photon exchange of the
incident electron with the target nucleus. In reality the dynamical picture of this process
can be strongly affected by many delicate effects. Many of them are important at low energies,
but in very specific cases they can manifest themselves at high energies too.

\subsection{Screening of the nucleus charge}

In an experimental setup, the target commonly consists of atoms, i.e. the initial electron
not only scatters off the nucleus with charge $Z e$ but it feels the influence of the
electron cloud around it. Roughly speaking, this leads to an effective decrease of
the target electric charge which is usually called {\it screening}.
In order to take this effect into account, one usually uses the atomic form factor $F(q^2)$ in the form
\eq{
	\frac{1}{q^2}
	\quad\longrightarrow\quad
	\frac{1 - F(q^2)}{q^2}
	=
	\sum_{i=1}^3 \frac{\alpha_i}{\beta_i^2 - q^2}.
	\label{eq.Screening}
}
Here we use the parametrization of this form factor from
the Moli{\`e}re approach \cite{Moliere:1947zza}
based on the Thomas--Fermi screening function.
The values of the parameters in (\ref{eq.Screening}) are the following:
\eq{
	\alpha_1 &= 0.1,
	\quad
	&\alpha_2 &= 0.55,
	\quad
	&\alpha_3 &= 0.35,
	\\
	b_1 &= 6.0,
	\quad
	&b_2 &= 1.2,
	\quad
	&b_3 &= 0.3,
}
with $\beta_i = \br{Z^{1/3}/121} m_e b_i$.
This approach is rather rough. For example, some important details of
the atomic structure can lead to sizeable effects \cite{Bondarenco:2023zoz}.
And then one should better use more elaborated models for atomic form factors,
for example, as in the paper \cite{Lobato:mq5024}.
But we limit ourselves to the Moli{\`e}re approach since
we mostly consider the low-energy regime of scattering
($E_1 \sim \mbox{few}~m_e$).

\subsection{Coulomb corrections}

Another effect is the {\it Coulomb corrections} that appear in the kinematic
regime when the electron moves with a small relative velocity $v$ with respect to the atom, and a new
important parameter $Z e^2/\hbar v$ appears. In that case, multiple scattering
can contribute significantly changing the physical picture essentially.
This specific mechanism cannot be taken into account in the
general case within quantum field theory.
It requires the use of precise solutions of the Dirac equation in the external field.
This was done in some specific cases, see, for example, Refs.~\cite{Tarasov:2011sa, Bondarenco:2022qyt}.
In our case, we neglect the effect of Coulomb corrections
by considering the electron as being far from the reaction threshold
where the relative velocity is considered to be large.

\subsection{Landau--Pomeranchuk--Migdal effect and others}

There are other effects which influence the bremsstrahlung process.
One of them is the well known Landau--Pomeranchuk--Migdal effect \cite{Landau:1953um,Migdal:1956tc},
which leads to a strong suppression of the cross section at small scattering angles.
The basic explanation of this effect involves negative interference in the forward region of two contributions coming from
two diagrams presented in Fig.~\ref{fig.Diagrams} (see the minus sign in one of two terms in (\ref{eq.OperatorOdefinition})).

For a comprehensive review of this effect and other processes, which influence the
cross section of bremsstrahlung in the medium of thick targets, one can see Ref.~\cite{Klein:1998du}.

\section{Comparison with previous results}
\label{sec.Comparison}

The work of Olsen and Maximon~\cite{Olsen:1959zz} is widely used to correct
experimental results for radiative effects. The calculation
includes polarization transer observables.
Further on, we will refer to this paper with the notation ``OM''.

First, we compare our calculation of the unpolarized cross section from equation (\ref{eq.def.SigmaI})
with the results reported in Fig.~1 or Ref.~\cite{Olsen:1959zz},
for the differential cross section $d\sigma/d\omega \, d\xi$.
The following relation holds:
\eq{
	\frac{d\sigma}{d\omega ~ d\xi}
	=
	J_\xi~ \frac{d\sigma}{d\omega ~dC_{1k}}.
	\label{eq.OMtoOurs}
}
where $\xi = 1/(1+u^2)$ and $u = \br{\brm{\vv{p_1}}/m_e} \sin \theta_{1k}$.
The quantity $J_\xi$ in (\ref{eq.OMtoOurs}) is the Jacobian that relates the
varibles ($C_{1k} \to \xi$) and has the form
\eq{
	J_\xi = \frac{dC_{1k}}{d\xi} = \frac{m_e^2}
	{2 \brm{\vv{p_1}} \xi^2 \sqrt{\brm{\vv{p_1}}^2 - m_e^2 \br{\frac{1}{\xi} - 1}}} \, .
}
Moreover, the cross section in Fig.~1 of Ref.~\cite{Olsen:1959zz}
is made dimensionless by using extra factor:
\eq{
	\phi = \br{e Z}^2 \frac{e^4}{m_e}.
}
\begin{figure}
	\centering
    \includegraphics[width=0.48\textwidth]{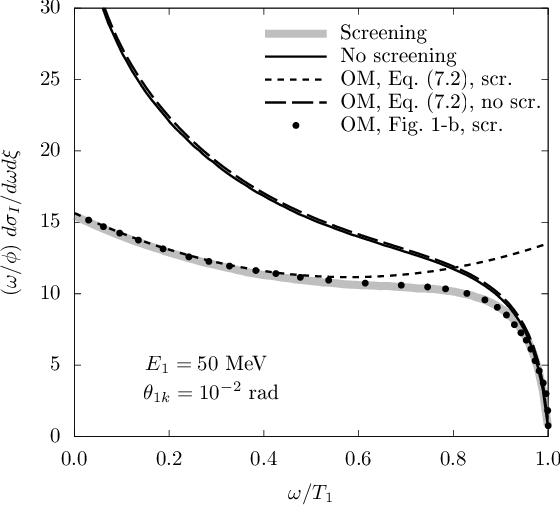}
	\caption{
    	The unpolarized cross section for scattering of
    	$50~\MeV$ electrons on lead at the photon emission
    	angle $\theta_{1k} = 10^{-2}~\rad$.
    }
    \label{fig.OM59Fig1}
\end{figure}
Thus, in Fig.~\ref{fig.OM59Fig1} we reproduce Fig.~1 of \cite{Olsen:1959zz}
where $T_1 \equiv E_1 - m_e$ is the kinetic energy of the incident electron.
The results of our exact calculation are presented as solid lines:
the solid black line is for the cross section without
effects of nucleus potential screening while the grey thick line is the same cross section including
the effects of screening (following the modification presented in (\ref{eq.Screening})).
It is clear that the screened cross section is smaller since the screening leads
to an effective decrease of the electric charge of the scattering center.
The dashed lines correspond to Eq.~(7.2) of Ref.~\cite{Olsen:1959zz}.
The long dashed lines correspond to no screening and the short dashed lines are for the screened cross section.
One can see that the present calculation coincides with the OM curve without screening while
the agreement with the screened curve takes place only for small energy photons ($\omega/T_1 < 0.5$).
This is the first problem we want to point out here: Eq.~(7.2) of paper \cite{Olsen:1959zz}
does not work for hard photon emission. One can notice that the dotted line taken from Fig.~1
in \cite{Olsen:1959zz} (black circles in Fig.~\ref{fig.OM59Fig1}) coincides with the
curve for the present calculation and it does not agree with Eq.~(7.2) of the same paper \cite{Olsen:1959zz}.

\begin{figure}
	\centering
    \includegraphics[width=0.48\textwidth]{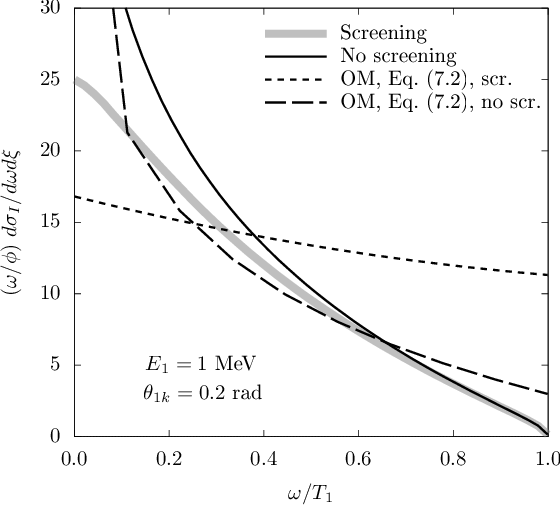}
	\caption{
    	The unpolarized cross section of
    	$1~\MeV$ electron scattering
    	in lead at the photon emission
    	angle $\theta_{1k} = 0.2~\rad$.
    }
    \label{fig.OM59Fig1a}
\end{figure}

Good agreement of the OM-result with the present calculation
occurs only for large electron energies and small scattering angles. In order to show this, we redraw
Fig.~\ref{fig.OM59Fig1} for small energy of the electron beam $E_1 = 1~\MeV$ and for a rather large photon
emission angle $\theta_{1k} = 0.2~\rad$ (see Fig.~\ref{fig.OM59Fig1a}).
One can see a large discrepancy between the OM results and the present calculation
at this electron energy and photon emission angle.

Next we consider the polarized observables. In Refs.~\cite{Olsen:1959zz, Haug:2004gp}
the Stokes parameter $P_1$ is named
{\it linear polarization} and $P_3$ stands for {\it circular polarization}.

\begin{figure}
	\centering
	\includegraphics[width=0.48\textwidth]{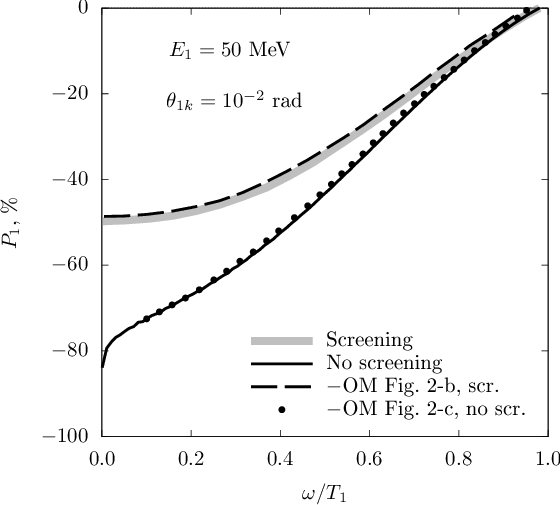}
    \caption{
    	Linear polarization of bremsstrahlung, $P_1$,
    	of $50~\MeV$ electrons
    	in lead at the photon emission
    	angle $\theta_{1k} = 10^{-2}~\rad$
    	(we integrate over $d\Omega_{\vv{p_3}}$ and $d\varphi_k$),
    	comparing with Fig.~2 of \cite{Olsen:1959zz}.
    }
    \label{fig.OM59Fig2}
\end{figure}

In Fig.~\ref{fig.OM59Fig2}, we reproduce Fig.~2 of paper \cite{Olsen:1959zz} that
corresponds to the case when the initial electron is not polarized while
the final photon is linearly polarized (Stokes parameter $P_1$).
One can see that the screened and not-screened cases are well reproduced.
The only difference is the total sign of the effect:
we plot the OM-curves with the opposite sign with respect to what is presented
in paper \cite{Olsen:1959zz}.
The reason for this difference is the different definitions of the photon purely
polarized states $\varepsilon_{1,2}^\mu$:
we chose 3-vector $\vv{\varepsilon_1}$ to lay in the photon scattering plane and
3-vector $\vv{\varepsilon_2}$ is perpendicular to it while Olsen and Maximon choose
the opposite definition (see the text and formula before Eq.~(7.3) in Ref.~\cite{Olsen:1959zz}).

\begin{figure}
	\centering
	\includegraphics[width=0.48\textwidth]{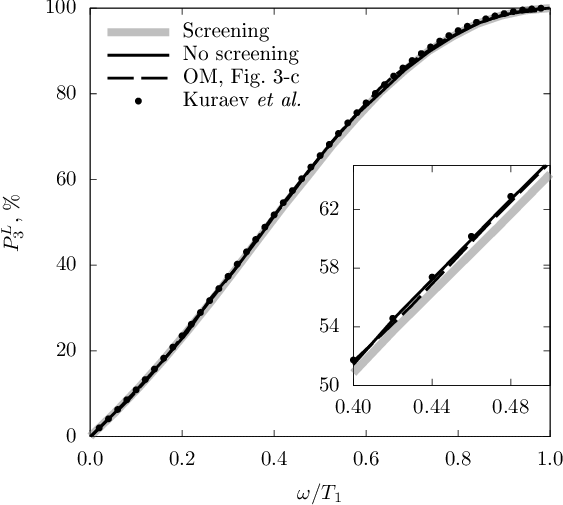}
    \caption{
    	Circular polarization of bremsstrahlung, $P_3^L$,
    	of longitudinally polarized ($\lambda_1 = +1$)
    	electrons with energy $E_1=50~\MeV$
    	in lead at the photon emission
    	angle $\theta_{1k} = 10^{-2}~\rad$
       	(we integrate over $d\Omega_{\vv{p_3}}$ and $d\varphi_k$).
    }
    \label{fig.OM59Fig3}
\end{figure}

In Fig.~\ref{fig.OM59Fig3}, we reproduce Fig.~3 of paper \cite{Olsen:1959zz}
that corresponds to the transfer of longitudinal
polarization of the initial electron with $\lambda_1 = +1$ to the
final photon probed to be circularly polarized (Stokes parameter $P_3$).
One can see good agreement of the present results (solid black and grey lines)
with the Olsen and Maximon results (black dashed lines).
In Fig.~\ref{fig.OM59Fig3} we also plot by black points the result of Prof.~E.A.~Kuraev {\it et al.}
(see Eq.~(46) in paper \cite{Kuraev:2009uy}) for the polarization transfer observable.
This result was obtained using the infinite momentum technique \cite{Baier:1980kx}
based on the Sudakov parametrization of vectors --- the light-cone decomposition \cite{Sudakov:1954sw}.
This technique applies to the peripheral kinematic regime when
$s \gg -q^2 \approx M_A^2 \gg m_e^2$. We see very good agreement of this calculation with the OM result
and with precise calculation. However, one should not expect this good agreement
for low energies and for kinematic regions where electron mass effects can be important.
We will show this issue further.

Moreover, all these calculations (by Olsen and Maximon and by Kuraev's group)
correspond to the unscreened case, but as one can see, the difference between the case of screening and
no screening for the observable $P_3^L$ is very small.

\begin{figure}
	\centering
	\includegraphics[width=0.48\textwidth]{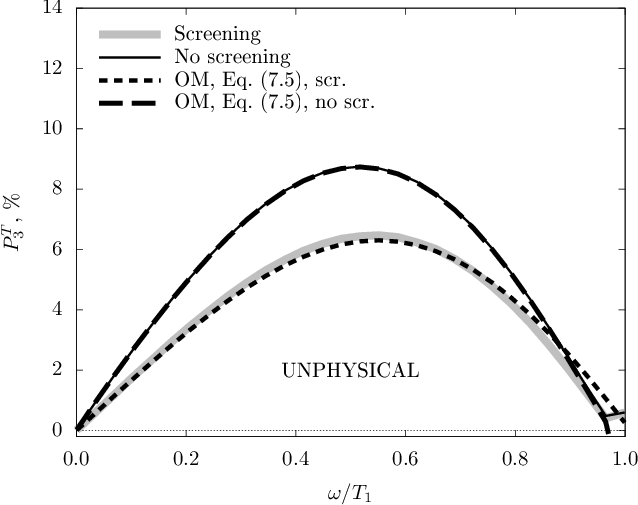}
    \caption{
    	Circular polarization of bremsstrahlung, $P_3^T$,
    	of transversely polarized electrons with energy $E_1=50~\MeV$
    	in lead at the photon emission angle $\theta_{1k} = 4.1~\mbox{mrad}$
    	(we integrate over $d\Omega_{\vv{p_3}}$ and $d\varphi_k$).
    }
    \label{fig.OM59Fig4}
\end{figure}

In Fig.~\ref{fig.OM59Fig4}, we reproduce Fig.~4 of paper \cite{Olsen:1959zz}
that corresponds to the transfer of transverse polarization from the initial electron
to the final photon probed to be circularly polarized (Stokes parameter $P_3$).
Again, one can see good agreement of the present calculation (solid black and grey lines) with the Olsen and Maximon
results (thick black dashed line).

The mark "UNPHYSICAL" on Fig.~\ref{fig.OM59Fig4} defines the unphysical setup,
which was chosen for Fig.~4 in Ref.~\cite{Olsen:1959zz}.
It is reproduced in the assumption that the photon is emitted in the plane contaning
an electron transverse polarization vector, i.e. $\varphi_k = \varphi_a$.
This does not correspond to the experimental situation:
in reality the photon is mostly emitted with some not zero azimuthal angle
with respect to a fixed electron polarization vector.
The absence of this comment in Ref.~\cite{Olsen:1959zz} might lead to erroneous understanding of the effect.
We comment on this situation in more detail in the next section.

\begin{table*}
\label{table.Comparison-OM-HN}
\caption{
	The comparison of our precise calculation with the results of Olsen--Maximon'59 (OM) and of Haug--Nakel'01 (HN).
   	We integrate not screened cross sections over $d\Omega_{\vv{p_3}}$ and $d\varphi_k$.
}
\centering
\begin{tabular}{|c||r|r|r||r|r|r||r|r|r|}
\hline
$\omega/T_1$ & \multicolumn{3}{c||}{$P_1$} & \multicolumn{3}{c||}{$P_3^L$} & \multicolumn{3}{c|}{$P_3^T$} \\
\hline
& ours & $-$OM & $-$HN & ours & OM & HN & ours & OM & HN \\
\hline
\multicolumn{10}{|c|}{$E_1 = 1$~$\MeV$, $\theta_{1k} = 0.25$~$\rad$} \\
\hline
$   0.1    $ & $             -0.1885$ & $             -0.2173$ & $             -0.1703$ & $              0.0490$ & $              0.0506$ & $              0.0473$ & $              0.0161$ & $              0.0104$ & $             -0.0059$ \\
$   0.2    $ & $             -0.1133$ & $             -0.1844$ & $             -0.0908$ & $              0.1007$ & $              0.1045$ & $              0.0964$ & $              0.0298$ & $              0.0176$ & $             -0.0185$ \\
$   0.3    $ & $             -0.0395$ & $             -0.1536$ & $             -0.0125$ & $              0.1548$ & $              0.1618$ & $              0.1468$ & $              0.0420$ & $              0.0221$ & $             -0.0374$ \\
$   0.4    $ & $              0.0384$ & $             -0.1217$ & $              0.0703$ & $              0.2104$ & $              0.2223$ & $              0.1977$ & $              0.0530$ & $              0.0233$ & $             -0.0631$ \\
$   0.5    $ & $              0.1232$ & $             -0.0870$ & $              0.1606$ & $              0.2666$ & $              0.2857$ & $              0.2477$ & $              0.0634$ & $              0.0208$ & $             -0.0964$ \\
$   0.6    $ & $              0.2176$ & $             -0.0483$ & $              0.2612$ & $              0.3218$ & $              0.3517$ & $              0.2947$ & $              0.0747$ & $              0.0139$ & $             -0.1381$ \\
$   0.7    $ & $              0.3258$ & $             -0.0039$ & $              0.3757$ & $              0.3740$ & $              0.4197$ & $              0.3355$ & $              0.0893$ & $              0.0013$ & $             -0.1889$ \\
$   0.8    $ & $              0.4549$ & $              0.0484$ & $              0.5100$ & $              0.4201$ & $              0.4892$ & $              0.3654$ & $              0.1123$ & $             -0.0185$ & $             -0.2494$ \\
$   0.9    $ & $              0.6190$ & $              0.1123$ & $              0.6740$ & $              0.4553$ & $              0.5594$ & $              0.3768$ & $              0.1539$ & $             -0.0484$ & $             -0.3192$ \\

\hline
\multicolumn{10}{|c|}{$E_1 = 3$~$\MeV$, $\theta_{1k} = 0.1$~$\rad$} \\
\hline
$   0.1    $ & $             -0.3794$ & $             -0.3838$ & $             -0.3726$ & $              0.0878$ & $              0.0883$ & $              0.0871$ & $              0.0199$ & $              0.0183$ & $              0.0139$ \\
$   0.2    $ & $             -0.3205$ & $             -0.3314$ & $             -0.3111$ & $              0.1855$ & $              0.1868$ & $              0.1838$ & $              0.0353$ & $              0.0316$ & $              0.0217$ \\
$   0.3    $ & $             -0.2630$ & $             -0.2798$ & $             -0.2504$ & $              0.2913$ & $              0.2940$ & $              0.2887$ & $              0.0467$ & $              0.0400$ & $              0.0238$ \\
$   0.4    $ & $             -0.2023$ & $             -0.2246$ & $             -0.1857$ & $              0.4025$ & $              0.4072$ & $              0.3991$ & $              0.0537$ & $              0.0429$ & $              0.0194$ \\
$   0.5    $ & $             -0.1379$ & $             -0.1646$ & $             -0.1161$ & $              0.5149$ & $              0.5225$ & $              0.5109$ & $              0.0562$ & $              0.0393$ & $              0.0076$ \\
$   0.6    $ & $             -0.0701$ & $             -0.0996$ & $             -0.0414$ & $              0.6235$ & $              0.6350$ & $              0.6185$ & $              0.0549$ & $              0.0285$ & $             -0.0124$ \\
$   0.7    $ & $              0.0005$ & $             -0.0296$ & $              0.0387$ & $              0.7229$ & $              0.7391$ & $              0.7143$ & $              0.0516$ & $              0.0099$ & $             -0.0413$ \\
$   0.8    $ & $              0.0750$ & $              0.0478$ & $              0.1284$ & $              0.8075$ & $              0.8295$ & $              0.7873$ & $              0.0519$ & $             -0.0182$ & $             -0.0805$ \\
$   0.9    $ & $              0.1650$ & $              0.1465$ & $              0.2460$ & $              0.8721$ & $              0.9018$ & $              0.8155$ & $              0.0741$ & $             -0.0629$ & $             -0.1335$ \\

\hline
\multicolumn{10}{|c|}{$E_1 = 10$~$\MeV$, $\theta_{1k} = 0.025$~$\rad$} \\
\hline
$   0.1    $ & $             -0.3262$ & $             -0.3273$ & $             -0.3266$ & $              0.1014$ & $              0.1014$ & $              0.1014$ & $              0.0243$ & $              0.0242$ & $              0.0238$ \\
$   0.2    $ & $             -0.2971$ & $             -0.2985$ & $             -0.2972$ & $              0.2151$ & $              0.2154$ & $              0.2153$ & $              0.0444$ & $              0.0441$ & $              0.0433$ \\
$   0.3    $ & $             -0.2657$ & $             -0.2672$ & $             -0.2653$ & $              0.3393$ & $              0.3396$ & $              0.3397$ & $              0.0597$ & $              0.0593$ & $              0.0579$ \\
$   0.4    $ & $             -0.2290$ & $             -0.2307$ & $             -0.2281$ & $              0.4696$ & $              0.4701$ & $              0.4708$ & $              0.0691$ & $              0.0682$ & $              0.0663$ \\
$   0.5    $ & $             -0.1867$ & $             -0.1886$ & $             -0.1853$ & $              0.5999$ & $              0.6005$ & $              0.6024$ & $              0.0711$ & $              0.0697$ & $              0.0671$ \\
$   0.6    $ & $             -0.1402$ & $             -0.1419$ & $             -0.1380$ & $              0.7224$ & $              0.7232$ & $              0.7268$ & $              0.0652$ & $              0.0629$ & $              0.0595$ \\
$   0.7    $ & $             -0.0914$ & $             -0.0926$ & $             -0.0880$ & $              0.8287$ & $              0.8298$ & $              0.8356$ & $              0.0518$ & $              0.0479$ & $              0.0433$ \\
$   0.8    $ & $             -0.0432$ & $             -0.0428$ & $             -0.0377$ & $              0.9120$ & $              0.9133$ & $              0.9207$ & $              0.0323$ & $              0.0253$ & $              0.0192$ \\
$   0.9    $ & $              0.0026$ & $              0.0070$ & $              0.0132$ & $              0.9673$ & $              0.9693$ & $              0.9745$ & $              0.0113$ & $             -0.0046$ & $             -0.0133$ \\

\hline
\multicolumn{10}{|c|}{$E_1 = 50$~$\MeV$, $\theta_{1k} = 0.0041$~$\rad$} \\
\hline
$   0.1    $ & $             -0.2503$ & $             -0.2513$ & $             -0.2512$ & $              0.1054$ & $              0.1055$ & $              0.1056$ & $              0.0260$ & $              0.0260$ & $              0.0260$ \\
$   0.2    $ & $             -0.2365$ & $             -0.2371$ & $             -0.2370$ & $              0.2233$ & $              0.2236$ & $              0.2237$ & $              0.0490$ & $              0.0491$ & $              0.0491$ \\
$   0.3    $ & $             -0.2196$ & $             -0.2194$ & $             -0.2194$ & $              0.3516$ & $              0.3521$ & $              0.3522$ & $              0.0680$ & $              0.0681$ & $              0.0681$ \\
$   0.4    $ & $             -0.1975$ & $             -0.1967$ & $             -0.1967$ & $              0.4855$ & $              0.4867$ & $              0.4869$ & $              0.0813$ & $              0.0815$ & $              0.0814$ \\
$   0.5    $ & $             -0.1685$ & $             -0.1685$ & $             -0.1684$ & $              0.6195$ & $              0.6206$ & $              0.6212$ & $              0.0871$ & $              0.0872$ & $              0.0872$ \\
$   0.6    $ & $             -0.1358$ & $             -0.1353$ & $             -0.1352$ & $              0.7436$ & $              0.7455$ & $              0.7465$ & $              0.0839$ & $              0.0840$ & $              0.0840$ \\
$   0.7    $ & $             -0.0988$ & $             -0.0985$ & $             -0.0984$ & $              0.8504$ & $              0.8521$ & $              0.8538$ & $              0.0714$ & $              0.0714$ & $              0.0713$ \\
$   0.8    $ & $             -0.0605$ & $             -0.0605$ & $             -0.0604$ & $              0.9316$ & $              0.9327$ & $              0.9350$ & $              0.0503$ & $              0.0501$ & $              0.0500$ \\
$   0.9    $ & $             -0.0244$ & $             -0.0242$ & $             -0.0241$ & $              0.9819$ & $              0.9822$ & $              0.9851$ & $              0.0231$ & $              0.0225$ & $              0.0223$ \\

\hline
\multicolumn{10}{|c|}{$E_1 = 100$~$\MeV$, $\theta_{1k} = 0.0025$~$\rad$} \\
\hline
$   0.1    $ & $             -0.3616$ & $             -0.3636$ & $             -0.3635$ & $              0.1063$ & $              0.1067$ & $              0.1067$ & $              0.0280$ & $              0.0281$ & $              0.0281$ \\
$   0.2    $ & $             -0.3398$ & $             -0.3438$ & $             -0.3438$ & $              0.2266$ & $              0.2269$ & $              0.2269$ & $              0.0531$ & $              0.0532$ & $              0.0532$ \\
$   0.3    $ & $             -0.3149$ & $             -0.3184$ & $             -0.3184$ & $              0.3583$ & $              0.3581$ & $              0.3582$ & $              0.0739$ & $              0.0739$ & $              0.0739$ \\
$   0.4    $ & $             -0.2838$ & $             -0.2851$ & $             -0.2851$ & $              0.4948$ & $              0.4952$ & $              0.4955$ & $              0.0882$ & $              0.0882$ & $              0.0882$ \\
$   0.5    $ & $             -0.2441$ & $             -0.2437$ & $             -0.2437$ & $              0.6307$ & $              0.6307$ & $              0.6311$ & $              0.0942$ & $              0.0943$ & $              0.0943$ \\
$   0.6    $ & $             -0.1956$ & $             -0.1954$ & $             -0.1953$ & $              0.7533$ & $              0.7554$ & $              0.7562$ & $              0.0904$ & $              0.0907$ & $              0.0907$ \\
$   0.7    $ & $             -0.1432$ & $             -0.1425$ & $             -0.1425$ & $              0.8584$ & $              0.8603$ & $              0.8614$ & $              0.0770$ & $              0.0771$ & $              0.0772$ \\
$   0.8    $ & $             -0.0888$ & $             -0.0886$ & $             -0.0886$ & $              0.9356$ & $              0.9378$ & $              0.9393$ & $              0.0548$ & $              0.0548$ & $              0.0549$ \\
$   0.9    $ & $             -0.0381$ & $             -0.0379$ & $             -0.0379$ & $              0.9829$ & $              0.9843$ & $              0.9862$ & $              0.0265$ & $              0.0264$ & $              0.0263$ \\

\hline
\end{tabular}
\end{table*}

In Table~\ref{table.Comparison-OM-HN}, we compare our calculations for
all three polarizations with two most widely used results ---
the results of Olsen and Maximon \cite{Olsen:1959zz} and the results of
Haug--Nakel \cite{Haug:2004gp}.
It is important to note that for the case of $P_1$ photon polarization
we again present the OM and HN results with the opposite sign because of different
definitions of the photon pure polarization states $\varepsilon_{1,2}^\mu$.
At moderate values of beam energy $E_1$, photon energies $\omega$ and
small emission angles $\theta_{1k}$ these approaches are in good agreement.
The agreement is getting worse at the boundaries of the kinematic region, especially
for large photon energies $\omega$.
This deviation is discussed in detail in the next section.

\section{Numerical analysis}
\label{sec.NumericalAnalysis}

In this section, a set of angular and energy distributions at moderately
high energies are presented to illustrate some general features of the
bremsstrahlung process. Specific
issues and the behaviour at low energies are discussed in the next section.

\begin{figure*}
	\centering
    \subfigure[]{\includegraphics[width=0.45\textwidth]{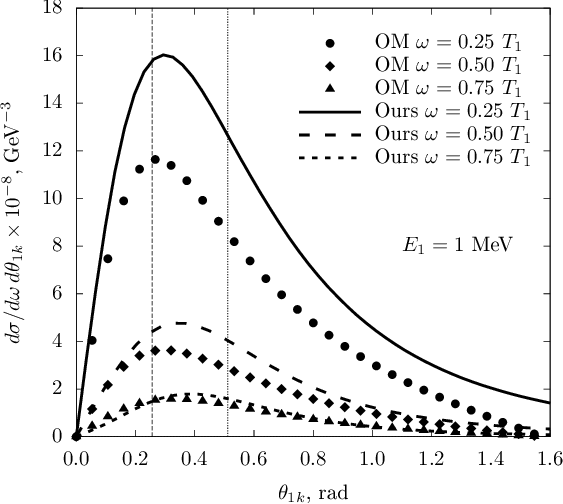}\label{fig.FigCSvsEmissionAngle1}}
	\hfill
    \subfigure[]{\includegraphics[width=0.45\textwidth]{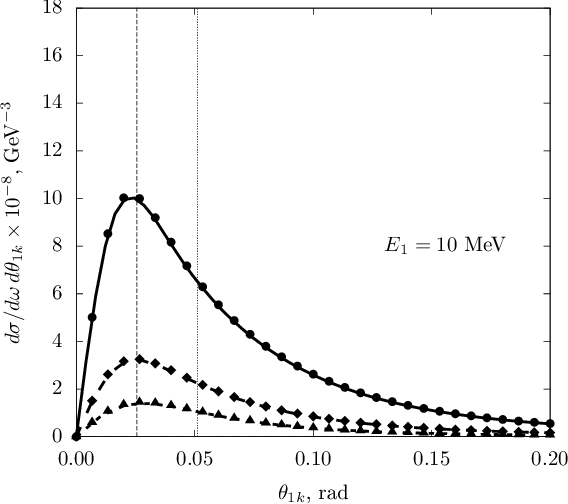}\label{fig.FigCSvsEmissionAngle2}}
    \\
    \subfigure[]{\includegraphics[width=0.45\textwidth]{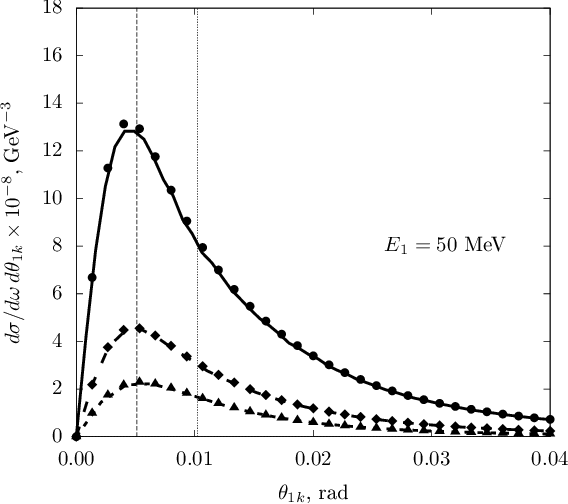}\label{fig.FigCSvsEmissionAngle3}}
	\hfill
    \subfigure[]{\includegraphics[width=0.45\textwidth]{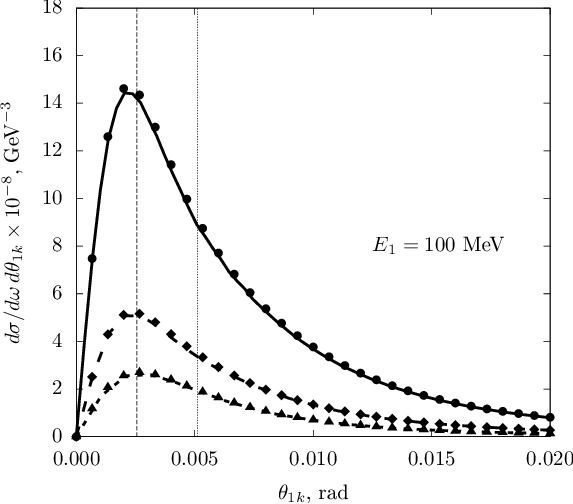}\label{fig.FigCSvsEmissionAngle4}}
    \caption{
    	The cross sections without screening as functions of the photon emission angle $\theta_{1k}$
    	for different initial unpolarized electron beam energies $E_1$ and for different photon energies $\omega$.
    	The vertical dotted line shows the position of the cone open angle $\theta_{1k} = m_e/E_1$,
    	while the vertical dashed line shows the position of the ``peak'' open angle $\theta_{1k} = m_e/2E_1$.
    	The curves are our exact results and the dots correspond to the calculations from paper \cite{Olsen:1959zz}.
		We integrate over $d\Omega_{\vv{p_3}}$ and $d\varphi_k$.
    }
    \label{fig.CSvsEmissionAngle}
\end{figure*}

The angular dependence of photon emission is illustrated in Fig.~\ref{fig.CSvsEmissionAngle}.
In shows the well-known peak-like behaviour along the initial electron beam.
Most of the contribution is concentrated inside the narrow cone with an upper angular limit
at $\theta_{1k} \sim m_e/E_1$. The position of this limit is shown in Fig.~\ref{fig.CSvsEmissionAngle}
by a vertical dotted line. Here we must stress that maximum of intensity lies closer
to the direction of the initial electron beam. The position of this
maximum slightly depends on the energy of the emitted photon $\omega$, but
approximately its position can be estimated as $\theta_{1k} \approx m_e/2 E_1$.
We show this value in Fig.~\ref{fig.CSvsEmissionAngle} by the vertical dashed lines.
Below we will plot the curves fixing the photon emission angle
$\theta_{1k}$ to this approximate maximum value, calling it ``{\it peak value}''.
The position of this ``peak value'' is indicated by the vertical dashed line in Fig.~\ref{fig.CSvsEmissionAngle}.
In this figure we also present the points which correspond to the results of Olsen and Maximon's
paper \cite{Olsen:1959zz}. We see that at large energies $E_1 \gg m_e$ the agreement of OM with
the present calculation is generally good except for the low-energy limit (say at $E_1 = 1~\MeV$),
where this agreement is absent
(see Fig.~\ref{fig.FigCSvsEmissionAngle1}). We see the same feature in other cases.

\begin{figure}
	\centering
	\includegraphics[width=0.48\textwidth]{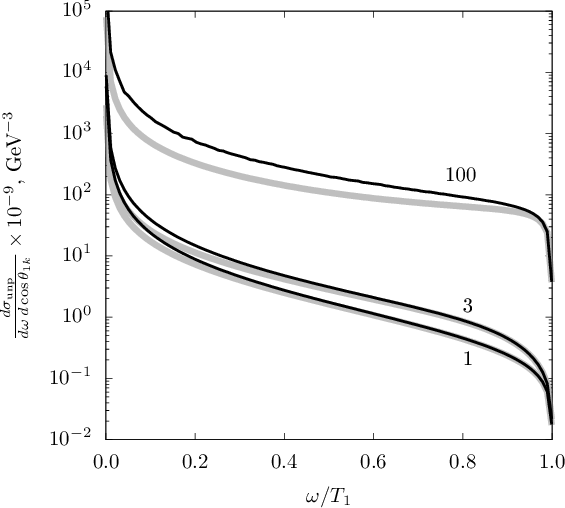}
    \caption{
    	The cross section as a function of emitted photon energy $\omega$
    	for different initial unpolarized electron beam energies $E_1$
    	(numbers near the curves in MeVs).
    	The photon emission angle is fixed at peaks (see the explanation in the text).
    	The black curves are without screening and the grey ones are with screening.
    	We integrate over $d\Omega_{\vv{p_3}}$ and $d\varphi_k$.
    }
    \label{fig.CSvsPhotonEnergy}
\end{figure}

In Fig.~\ref{fig.CSvsPhotonEnergy}, the unpolarized cross section as a function of emitted photon energy $\omega$
for different beam energies $E_1$ is presented.
All curves are plotted at peaks (i.e. $\theta_{1k} \approx m_e/2 E_1$).
Again one can see that the screened cross sections
(grey thick lines) systematically lie below the not-screened ones (black lines).

\begin{figure}
	\centering
	\includegraphics[width=0.48\textwidth]{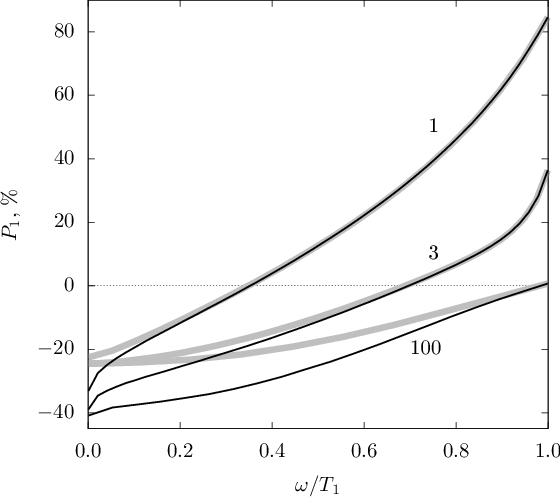}
    \caption{
    	The linear polarization of the emitted photon, $P_1$, versus its energy $\omega$
    	for the unpolarized initial electron beam with different energies $E_1$.
    	The notation is as in Fig.~\ref{fig.CSvsPhotonEnergy}.
    }
    \label{fig.P1vsPhotonEnergy}
\end{figure}

\begin{figure}
	\centering
	\includegraphics[width=0.48\textwidth]{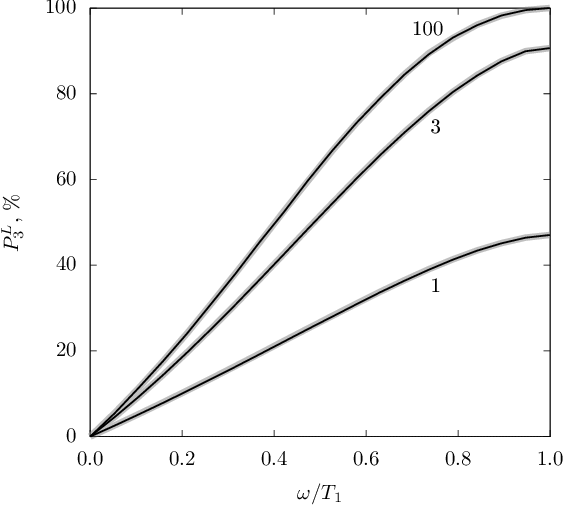}
    \caption{
    	The circular polarization of the emitted photon, $P_3^L$, versus its energy $\omega$
    	for the longitudinally polarized
    	($\lambda_1 = +1$) initial electron beam with different energies $E_1$.
    	The notation is as in Fig.~\ref{fig.CSvsPhotonEnergy}.
    }
    \label{fig.P3vsPhotonEnergy}
\end{figure}

In Fig.~\ref{fig.P1vsPhotonEnergy} and Fig.~\ref{fig.P3vsPhotonEnergy},
we show how the polarization of the final photon depends on
the photon energy $\omega$. One can see that these quantities weakly depend on
the screening effects.

\begin{figure*}
	\centering
    \subfigure[]{\hspace{-10mm}\includegraphics[width=0.45\textwidth]{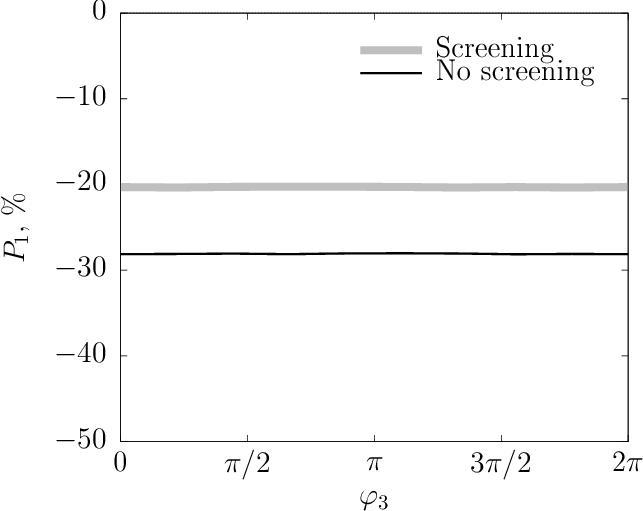}\label{fig.FigP1vsPhi3}}
	\hspace{15mm}
    \subfigure[]{\hspace{-10mm}\includegraphics[width=0.45\textwidth]{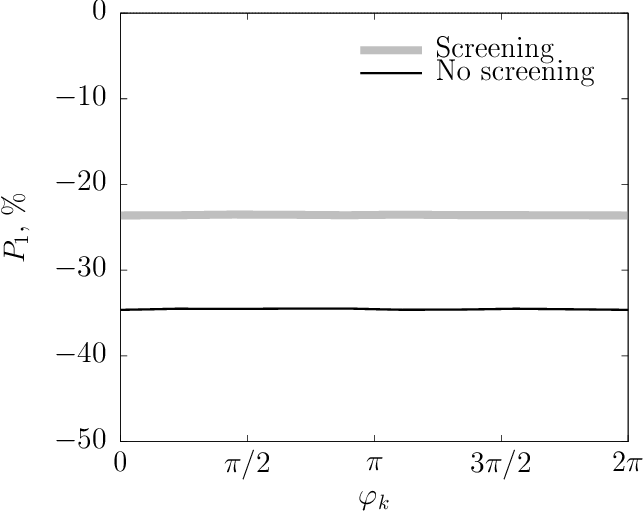}\label{fig.FigP1vsPhik}}
    \\
    \subfigure[]{\hspace{-10mm}\includegraphics[width=0.45\textwidth]{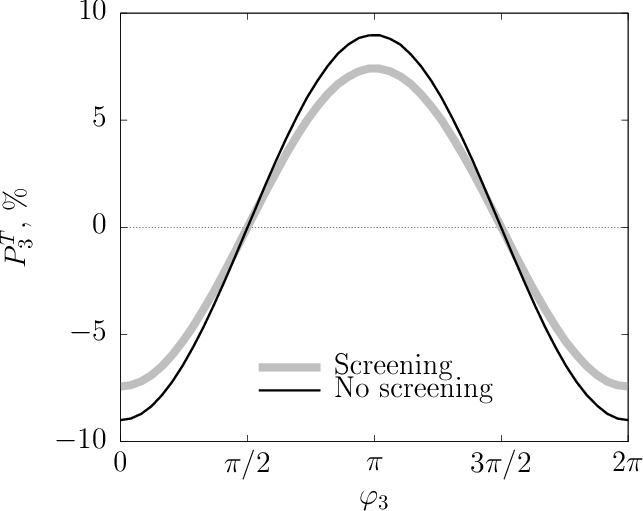}\label{fig.FigP3vsPhi3}}
	\hspace{15mm}
    \subfigure[]{\hspace{-10mm}\includegraphics[width=0.45\textwidth]{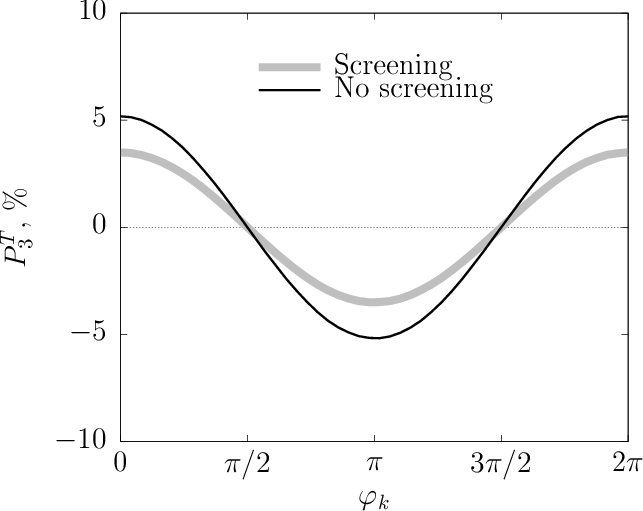}\label{fig.FigP3vsPhik}}
    \caption{
    	The polarization degrees of emitted photons with polarization $P_1$ and $P_3^T$ as functions of the
    	electron azimuthal emission angle $\varphi_3$ and the
    	photon azimuthal emission angle $\varphi_k$ for the fixed photon emission angle $\theta_{1k}$ at peak value.
    	The initial electron beam energy is $E_1 = 50~\MeV$ and photon energy is fixed at $\omega = 0.4 E_1$ (left plots)
    	and $\omega = 0.2 E_1$ (right plots).
    	We integrate over $dC_{13}$ and $d\varphi_k$ in (\ref{fig.FigP1vsPhi3}) and (\ref{fig.FigP3vsPhi3})
    	and over $dC_{13}$ and $d\varphi_3$ in (\ref{fig.FigP1vsPhik}) and (\ref{fig.FigP3vsPhik}).
    }
    \label{fig.P1andP3vsPhis}
\end{figure*}

Now let us investigate the azimuthal dependence of these polarization degrees.
These dependences for the $P_1$ polarization degree are
shown in Figs.~\ref{fig.FigP1vsPhi3} and \ref{fig.FigP1vsPhik}.
In the first plot in Fig.~\ref{fig.FigP1vsPhi3} you can see the dependence of the emitted photon
probed to be in a polarization state with the
$P_1$ Stokes parameter as a function of the scattered electron azimuthal angle $\varphi_3$
(integrated over $\varphi_k$) for electron beam energy fixed to $E_1 = 50~\MeV$,
and photon has also fixed energy $\omega = 0.2 E_1$.
While in Fig.~\ref{fig.FigP1vsPhik} this observable is plotted as a function of the emitted
photon azimuthal angle $\varphi_k$.
One can see that these values are independent of the azimuthal angles and are finite.

The azimuthal dependences of the polarization state $P_3$
are shown in Figs.~\ref{fig.FigP3vsPhi3} and \ref{fig.FigP3vsPhik},
which are generated by the transverse initial electron spin directed along the $x$-axis.
Here one can see why we marked the plot in Fig.~\ref{fig.OM59Fig4} as unphysical.
The photon is emitted at a definite azimuthal angle with respect to the initial beam
spin orientation, and if one integrates over this orientation,
then the integrated value $P_3^T$ vanishes. That is why the physical plot
in Fig.~\ref{fig.OM59Fig4} will be zero.

\begin{figure}
    \includegraphics[width=0.45\textwidth]{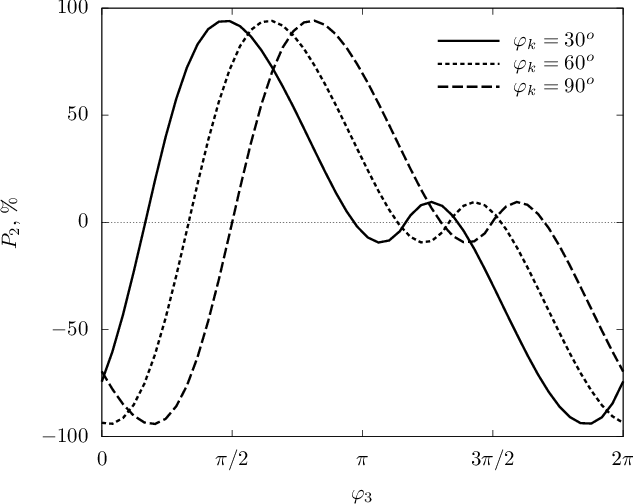}
    \caption{
    	The not screened polarization degrees of emitted photons with polarization $P_2$ as functions of the
    	electron azimuthal emission angle $\varphi_3$ (for fixed $\varphi_k$).
    	The photon emission angle $\theta_{1k}$ is fixed at peak value.
    	The initial electron beam energy is $E_1 = 50~\MeV$ and photon energy is fixed at $\omega = 0.2 E_1$.
    	We integrate over $dC_{13}$.
    }
    \label{fig.P2vsPhi3}
\end{figure}

As for the photon polarization $P_2$, in Ref.~\cite{Olsen:1959zz} it was not illustrated
because the authors supposed it to be zero.
It is true for an integrated observable, but
$P_2$ is not zero when both azimuthal angles are fixed. For example in Fig.~\ref{fig.P2vsPhi3} we present the
$P_2^T$ value as a function of the scattered electron azimuthal angle $\varphi_3$ with the fixed
photon azimuthal angle $\varphi_k$. It is clear that if one integrates over $\varphi_3$, then
the quantity $P_2^T$ becomes zero for each fixed $\varphi_k$. The same is true if one integrates
over $\varphi_k$ with the fixed $\varphi_3$.

\section{Issues and problems}
\label{sec.Issues}

In this section we illustrate the problems that arise with a naive exploitation
of the previous results. Sometimes they are hidden during the modelling of the
bremsstrahlung process and appear only in specific configurations and
settings:

\begin{enumerate}

	\item
	As it was already noticed in Section~\ref{sec.Comparison}, if one uses Eq.~(7.2) of Ref.~\cite{Olsen:1959zz}
	to evaluate the unpolarized cross section of the bremsstrahlung process with screening of the nucleus potential
	at large emitting photon energy $\omega$
	(say, for $\omega > 0.5 \, T_1$), then one can get wrong numbers even for high energies
	of an electron beam, like $E_1 = 50~\MeV$. This is obviously seen in Fig.~\ref{fig.OM59Fig1}.
    However, it is clear that the authors of Ref.~\cite{Olsen:1959zz} used a
	correct formula for plotting the curve on Fig.~1 of paper \cite{Olsen:1959zz}.
	That curve is presented in Fig.~\ref{fig.OM59Fig1} with black dots and it agrees with our result.
	So the usage of Eq.~(7.2) of Ref.~\cite{Olsen:1959zz} for hard photons is wrong.

	\item
	There is a systematic problem with an approximate result of Ref.~\cite{Olsen:1959zz}
	for the unpolarized cross section at low energies, for $E_1 \sim \mbox{few}~m_e$.
	Figure~\ref{fig.FigCSvsEmissionAngle1} shows a large discrepancy between the
	results from Ref.~\cite{Olsen:1959zz} and the present work.
	This difference can reach 30--40 percent even for bremsstrahlung photons
	with low energy ($\omega < 0.5 \, T_1$).

\begin{figure}
	\centering
	\includegraphics[width=0.48\textwidth]{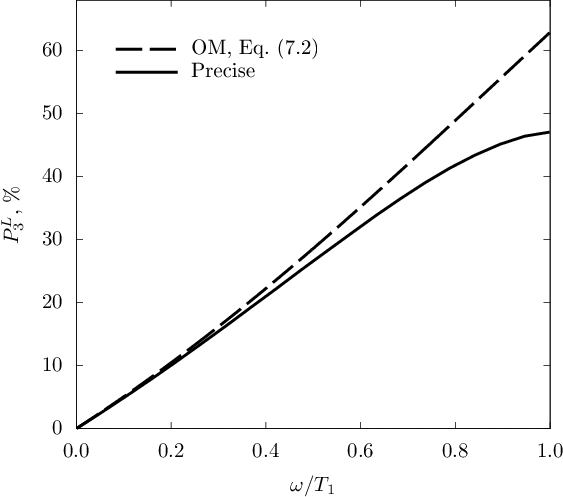}
    \caption{
    	The not screened circular polarization of the emitted photon, $P_3^L$, versus its energy $\omega$
    	for the initial longitudinally polarized
    	($\lambda_1 = +1$) electron beam with the energy $E_1 = 1~\MeV$
    	and peak emission angle.
    	The comparison with the OM result is given.
    	We integrate over $d\Omega_{\vv{p_3}}$ and $d\varphi_k$.
    }
    \label{fig.P3vsPhotonEnergyAtSmallEnergy}
\end{figure}

	\item
	Problems can be pointed out for polarization observables as well.
	In Fig.~\ref{fig.P3vsPhotonEnergyAtSmallEnergy} we redraw Fig.~\ref{fig.OM59Fig1}
	for a low electron beam energy $E_1 = 1~\MeV$ and for a rather large photon
	emission angle $\theta_{1k} = m_e/2E_1 = 0.2~\rad$.
	One can see a difference between the two results.
  	This difference increases when the initial electron energy $E_1$ decreases.
	The reason is that the OM calculation systematically
	neglects the electron mass and the nucleus recoil, which are very important in that case.

\begin{figure}
	\centering
	\includegraphics[width=0.48\textwidth]{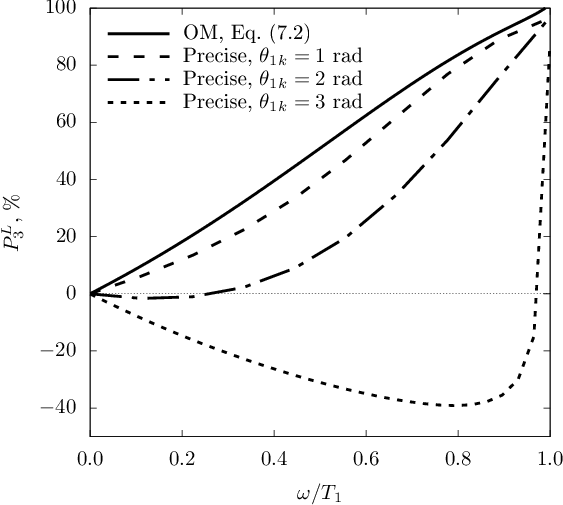}
    \caption{
    	The not screened circular polarization of the emitted photon, $P_3^L$, versus its energy $\omega$
    	for the initial longitudinally polarized
    	($\lambda_1 = +1$) electron beam with energy $E_1 = 3~\MeV$.
    	The emission angles are large ($\theta_{1k} = 1, 2, 3~\rad$).
    	We integrate over $d\Omega_{\vv{p_3}}$ and $d\varphi_k$.
    }
    \label{fig.P3vsPhotonEnergyAtLargeAngles}
\end{figure}

	\item
	Moreover, one can observe the noticeable non-linearity
	of $P_3$ of the bremsstrahlung photon with
	respect to its energy. See Fig.~\ref{fig.OM59Fig3} for a high beam energy $E_1 = 50~\MeV$
	and Fig.~\ref{fig.P3vsPhotonEnergyAtLargeAngles} for a low beam energy $E_1 = 3~\MeV$.
	It is seen from Fig.~\ref{fig.P3vsPhotonEnergyAtLargeAngles} that
	the bigger the photon emission angle $\theta_{1k}$, the stronger
	the distortion of the linear dependence.
    The Olsen and Maximon result does not depend on the angle $\theta_{1k}$
    and is calculated in the small angle approximation.
	Why is this issue important?
	In many papers the linear approximation is often used for estimation.
	See, for example, Ref.~\cite{Potylitsyn:1997wn}, where
	this linear approximation is used for a brief estimation of
	the bremsstrahlung photon polarization.
	This approximation works well at low photon energies and
	small emission angles, but when its energy tends to the upper kinematic limit,
	or its emission angle becomes large, this approximation fails, sometimes very dramatically.
	
\begin{figure}
	\centering
	\includegraphics[width=0.48\textwidth]{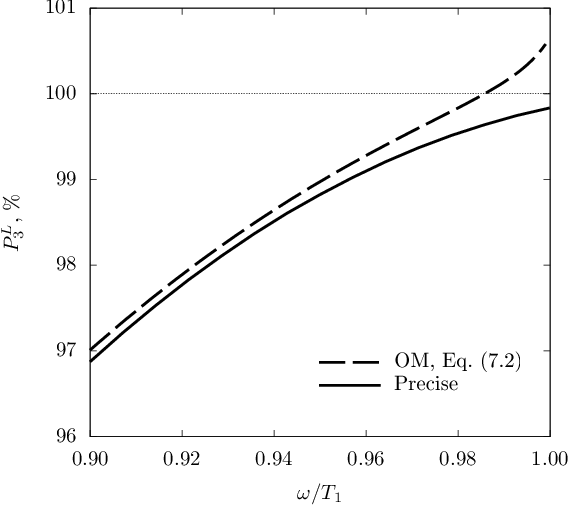}
    \caption{
    	The circular polarization of the emitted photon, $P_3$, versus its energy $\omega$
    	at the upper limit
    	for the initial longitudinally polarized
    	($\lambda_1 = +1$) electron beam with energy $E_1 = 10~\MeV$
    	and the emission angle is $\theta_{1k} = 1~\rad$.
    	We integrate over $d\Omega_{\vv{p_3}}$ and $d\varphi_k$.
    }
    \label{fig.P3vsPhotonEnergyAtUpperCorner}
\end{figure}

	\item
	And the last but not least issue is related to unphysical behaviour of
	the approximated formulae of the Olsen and Maximon paper \cite{Olsen:1959zz}
	near the boundaries of the physically allowed region.
	This is shown in Fig.~\ref{fig.P3vsPhotonEnergyAtUpperCorner}, where
	the circular polarization of the emitted photon, $P_3$, is plotted as a function of
	its energy $\omega$ for a rather large beam energy $E_1 = 10~\MeV$ and a large emission
	angle $\theta_{1k} = 1~\rad$. One can see that the OM result becomes unphysical
	and exceeds 100\% value of polarization for hard photons when they receive
	almost all the energy of the initial electron. The reason for this situation
	is again the systematical neglect of the electron mass $m_e$, which allows
	to simplify the mathematical expressions but narrows the region of their applicability.
	This may be occasionally useful, but one has to be aware of the
	limits of the used approximations.

\end{enumerate}

In the present experiments the Monte Carlo generators are essential to the data analysis.
The kinematics in the whole phase volume of final particles needs to be known and
reconstructed.
Thus, the approach of infinite momentum technique, which was used in the calculation of
Kuraev {\it et al.} in Ref.~\cite{Kuraev:2009uy}, is not suitable for this task
since it is limited to the peripheral kinematic regime when $s \gg -q^2 \approx M_A^2 \gg m_e^2$.

Caution must be taken when using the well-known book on bremsstrahlung
by Haug--Nakel \cite{Haug:2004gp} since many results presented there are approximate and
mostly follow the approach of Ref.~\cite{Bethe:1934za}. The comparison of their results
with the present calculation also shows the shortcomings listed above (see Table~\ref{table.Comparison-OM-HN}).

For convenience of a future Monte Carlo generator developer we present
explicit formulae for the precise cross sections in Appendix~\ref{sec.ExplicitCrossSections}.

\section{Conclusion}

In this work we have calculated the unpolarized
cross section and polarization observables for the bremsstrahlung of the electron in
scattering off the nucleus (Bethe--Heitler process).
The calculation was carried out in the first order of perturbation theory within QED.

We systematically keep the mass of electron $m_e$ and take into account the effects of
nucleus recoil. We do not consider the effects of emission from the nucleus keeping
in mind that for relatively low energies this contribution remains negligible.
In a separate section we discuss the aspects which might extend our first order calculation
in order to take into account more delicate effects such as nucleus field screening or Coulomb corrections
near the threshold.

We have performed a broad comparison of our calculation with the known ones, which are
often used to estimate radiative effects in different experimental setups.
Section \ref{sec.Issues} contains a list of issues which one must keep in mind during the
application of the known formulae in everyday practice. We show that the systematic application of
precise formulae solves all the problems related to the
approximations used in the previous papers and should be recommended for use,
in particular for developing stable Monte-Carlo generators.

The aforementioned precise formulae of the cross sections for unpolarized and polarized bremsstrahlung
derived in
this work are presented in Appendix~\ref{sec.ExplicitCrossSections}
and give us an appropriate description of the Bethe--Heitler process at relatively low energies.
This calculation is important for future polarization experiments and
for developing a PEPPo-like polarized positron source.

\begin{acknowledgments}
We are grateful to Eric Voutier for turning our attention to this problem, physical motivation of the task
and for valuable contribution at the early stage of the work.
The authors are grateful to Egle Tomasi-Gustafsson for fruitful discussions and help with the manuscript.
This project has received funding from the European Union's Horizon 2020 research and innovation
program under grant agreement No.~824093.
V.Z. work was supported partly by the Convergence--2025 Research Program
of Republic of Belarus (Microscopic World and Universe Subprogram).
\end{acknowledgments}

\appendix

\section{Photon pure polarization states}
\label{sec.PhotonPolarizationVectors}

In this section we construct two independent polarization states of the photon emitted in a specific direction.

We follow the basic idea and notation from paper \cite{McMaster:1954}.
Polarization of the photon is usually described by the direction of vibration of
the electric vector $\vv{E}$, which is perpendicular to the direction the photon propagation.
This electric vector can be considered as a linear combination of two orthogonal unit vectors
$\vv{E_1}$ and $\vv{E_2}$ as follows:
\eq{
	\vv{E} = a_1 \vv{E_1} + a_2 \vv{E_2},
}
where $a_{1,2}$ are in general complex scalars which describe the amplitude
and phase of two vibrations. From these two expansion coefficients we can form
a $2 \times 2$ density matrix \cite{McMaster:1961}:
\eq{
	\rho
	=
	\br{
		\begin{matrix}
		 a_1 a_1^* & a_1 a_2^*	\\
		 a_2 a_1^* & a_2 a_2^*	\\
		\end{matrix}
	}
	=
	\frac{1}{2}
	\br{
		\begin{matrix}
		 1 + P_1 & P_2 + i P_3	\\
		 P_2 - i P_3 & 1 - P_1	\\
		\end{matrix}
	},
}
which is normalized to unity and is parameterized in terms of the Stokes parameters $P_i$.
So to calculate the polarization state of the emitted photon, we must construct two
orthogonal vectors $\vv{E_{1,2}}$ and calculate the projection coefficients $a_i$.

\begin{figure}
	\centering
    \includegraphics[width=0.3\textwidth]{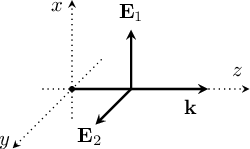}
    \caption{
    	Initial disposition of the pure polarization photon states.
    }
    \label{fig.PolarizationVectors1}
\end{figure}

\begin{figure}
	\centering
    \includegraphics[width=0.3\textwidth]{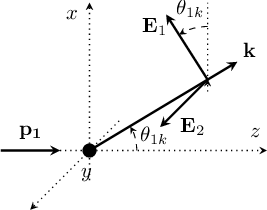}
    \caption{
    	The pure polarization photon states after polar rotation.
    }
    \label{fig.PolarizationVectors2}
\end{figure}

\begin{figure}
	\centering
    \includegraphics[width=0.19\textwidth]{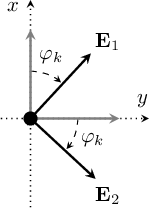}
    \caption{
    	The pure polarization photon states after azimuthal rotation.
    }
    \label{fig.PolarizationVectors3}
\end{figure}

First, we select the photon to go along the $z$-axis:
\eq{
	k^\mu = \br{ k^0, k^x, k^y, k^z} = \br{\omega, 0, 0, \omega}.
	\nn
}
Then as a pure polarization vectors $\vv{E}_{1,2}$ one can choose two
unit vectors along $x$ and $y$ axes (see Fig.~\ref{fig.PolarizationVectors1}), i.e. construct two 4-vectors
$\varepsilon_{1,2}^\mu = \br{0, \vv{E}_{1,2}}$:
\eq{
	\varepsilon_1^\mu &= \br{ e_1^0, e_1^x, e_1^y, e_1^z} = \br{ 0, 1, 0, 0 }, \nn\\
	\varepsilon_2^\mu &= \br{ e_2^0, e_2^x, e_2^y, e_2^z} = \br{ 0, 0, 1, 0 }. \nn
}

Next turn the photon momentum $\vv{k}$ and polarization vectors $\vv{E}_{1,2}$
by the angle $\theta_{1k}$ around the $y$-axis clockwise, as it is presented
in Fig.~\ref{fig.PolarizationVectors2}. The vector $\vv{E}_1$ lies in the $xz$-plane and
is perpendicular to the $y$-axis; thus,
\eq{
	\varepsilon_1^\mu &=
	\br{ 0, C_{1k}, 0, -S_{1k}}, \nn
}
while the vector $\vv{E}_2$ remains unchanged.
The photon momentum $k^\mu$ then becomes equal to:
\eq{
	k^\mu = \omega \br{1, S_{1k}, 0, C_{1k}}.
}

And finally, we need to rotate all these vectors around the $z$-axis by the azimuthal angle $\varphi_k$ clockwise
(see Fig.~\ref{fig.PolarizationVectors3} for details, the $z$-axis is directed from top to bottom of the figure).
Then the $z$-components of all the vectors remain the same and the rotation on the $x$ and $y$ components is applied:
\eq{
	\varepsilon_1^\mu &= \br{ 0, C_{1k} \cos\varphi_k, C_{1k} \sin\varphi_k, -S_{1k}}, \label{eq.E1}
	\\
	\varepsilon_2^\mu &= \br{ 0, -\sin\varphi_k, \cos\varphi_k, 0}, \label{eq.E2}
	\\
	k^\mu &= \omega \br{ 1, S_{1k} \cos\varphi_k, S_{1k} \sin\varphi_k, C_{1k}}. \label{eq.k}
}
It is easy to check that these vectors satisfy all the necessary conditions:
\eq{
	\br{\varepsilon_1 \varepsilon_2} = 0,
	\hspace{5mm}
	\varepsilon_i^2 = -1,
	\hspace{5mm}
	\br{\varepsilon_i k} = 0,
	\hspace{5mm}
	i=1,2.
}
We use these vectors in our numerical estimations.

\begin{table}
\caption{Components of 4-momenta}
\label{table.VectorComponents}
\begin{tabular}{c|cccc}
\hline
      &  $0$   &  $x$  &  $y$  &   $z$   \\
 \hline
 $p_1$  & $E_1$  &  $0$  &  $0$  & $\brm{\vv{p_1}}$  \\
 $p_2$  & $M_A$  &  $0$  &  $0$  & $ 0 $  \\
 $p_3$  & $E_3$  &  $\brm{\vv{p_3}} S_{13} \cos\varphi_3$  &
 $\brm{\vv{p_3}} S_{13} \sin\varphi_3$ & $\brm{\vv{p_3}} C_{13}$  \\
 $k$  & $\omega$  &  $\omega S_{1k} \cos\varphi_k$  &  $\omega S_{1k} \sin\varphi_k$  & $\omega C_{1k}$  \\
\hline
 $a_1^{\|}$    & $\lambda_1 \brm{\vv{p_1}}/m_e$  &  $0$  &  $0$  & $\lambda_1 E_1/m_e$  \\
 $a_1^{\perp}$ & $0$  &  $\cos\varphi_a$  &  $\sin\varphi_a$  & $0$  \\
\hline
 $\varepsilon_1$ & $0$  &  $C_{1k} \cos \varphi_k$  &  $C_{1k} \sin \varphi_k$  & $-S_{1k}$  \\
 $\varepsilon_2$ & $0$  &  $-\sin \varphi_k$  &  $\cos \varphi_k$  & $0$  \\
\hline
\end{tabular}
\end{table}

\section{Explicit expressions for cross sections}
\label{sec.ExplicitCrossSections}

In this section we present the explicit expressions for the unpolarized and
polarized cross sections of the bremsstrahlung process.
The figures plotted above (except for Figs.~\ref{fig.P1andP3vsPhis} and \ref{fig.P2vsPhi3}) contain cross sections as functions of the photon energy $\omega$
for fixed values of the cosine of the emission angle $C_{1k}$ or vice versa:
\eq{
	\frac{d\sigma_i}{d\omega dC_{1k}}
	&=
	\frac{\alpha^3 Z^2}{2^6 \pi^2 F}
	\int\limits_{-1}^1 dC_{13}
	\int\limits_{0}^{2\pi}d\varphi_3
	\int\limits_{0}^{2\pi}d\varphi_k
	\,
	\frac{\omega F_b}{(q^2)^2}
	\,
	S_i,
}
where index $i$ defines the type of the cross section and the
quantity $S_i$ is expressed via $S_{ij}$ from (\ref{eq.SbDefinition}) as:
\eq{
	S_I	    &= S_{11} + S_{22} = 16 \, {\cal S}_I,    		\label{eq.def.calSI}\\
	S_{P_1} &= S_{11} - S_{22} = 8  \, {\cal S}_{P_1},		\label{eq.def.calSP1}\\
	S_{P_2} &= S_{12} + S_{21} = 8  \, {\cal S}_{P_2},		\label{eq.def.calSP2}\\
	S_{P_3} &= i\br{ S_{21} - S_{12}} = 4 \, {\cal S}_{P_3}.\label{eq.def.calSP3}
}

\subsection{Unpolarized cross section}

Here we present the quantity ${\cal S}_I$ from (\ref{eq.def.calSI})
which is for the unpolarized cross section. It has the following structure:
\eq{
&
{\cal S}_I =
\frac{A_1}{w_1 w_3}
+
\frac{A_2}{w_1}
+
\frac{A_3}{w_3}
+
2 m_e^2 \br{
\frac{A_4}{w_1^2}
+
\frac{A_5}{w_3^2}
},
}
where the coefficients $A_i$ have the form:
\eq{
A_1 =&
- 4 M_A^2 p_{13} \br{ p_{13} - 4 m_e^2 } -
\nn\\&
- p_{13} \br{ w_2 \br{ p_{34} - p_{14} } - w_4 \br{ p_{12} - p_{23} } } +
\nn\\&
+ 2 p_{13} \br{ p_{12} p_{34} + p_{23} p_{14} }
- 4 m_e^2 \br{ w_2 w_4 + p_{13} p_{24} },
\nn\\
A_2 =&
w_2 p_{34} + w_4 p_{23} + 2 M_A^2 \br{ 2 p_{13} - w_3 } -
\nn\\&
- p_{12} p_{34} - 2 p_{12} p_{14} - p_{23} p_{14}
+ 4 m_e^2 M_A^2,
\nn\\
A_3 =&
w_2 p_{14} + w_4 p_{12} - 2 M_A^2 \br{ 2 p_{13} + w_1 } +
\nn\\&
+ p_{12} p_{34} + 2 p_{23} p_{34} + p_{23} p_{14}
- 4 m_e^2 M_A^2,
\nn\\
A_4 =&
 w_2 p_{34} + w_4 p_{23} - p_{12} p_{34} - p_{14} p_{23} +
\nn\\&
+ 2 M_A^2 \br{ p_{13} - w_3 }
+ 2 m_e^2 \br{ p_{24} - 4 M_A^2 },
\nn\\
A_5 =&
- w_2 p_{14} - w_4 p_{12} - p_{12} p_{34} - p_{23} p_{14} +
\nn\\ &
+ 2 M_A^2 \br{ p_{13} + w_1 }
+ 2 m_e^2 \br{ p_{24} - 4 M_A^2 }.
\nn
\label{expr}
}
Here we used a few radiative invariants $ w_i \equiv 2\br{k p_i}$
and scalar products $p_{ij} \equiv 2 \br{p_i p_j}$.
These invariants are calculated using the explicit components
in the laboratory frame presented in Table~\ref{table.VectorComponents}.

\subsection{Linear polarization}

Here we present the quantities ${\cal S}_{P_{1,2}}$ from (\ref{eq.def.calSP1}) and (\ref{eq.def.calSP2})
which are the cross sections of emission of the linearly polarized photon:
\eq{
&{\cal S}_{P_1} =
\frac{A_6}{w_1 w_3}
+
\frac{A_7}{w_1}
+
\frac{A_8}{w_3}
+
\frac{A_9}{w_1^2}
+
\frac{A_{10}}{w_3^2}
+
A_{11},
\\
&{\cal S}_{P_2} =
\frac{A_{12}}{w_1 w_3}
+
\frac{A_{13}}{w_1}
+
\frac{A_{14}}{w_3}
+
\frac{A_{15}}{w_1^2}
+
\frac{A_{16}}{w_3^2}
+
A_{17},
}
with the quantities $A_i$ equal to:
\eq{
A_6 =&
\br{ x_1 x_3 - y_1 y_3 }
\left(
2\br{ w_2 w_4 - p_{12} p_{34} - p_{23} p_{14} } -
\right.
\nn\\&
- w_2 \br{ p_{14} - p_{34} }
- w_4 \br{ p_{12} - p_{23} } +
\nn\\&
\left.
	+ 4 \br{ M_A^2 p_{13} + m_e^2 p_{24} - 4 m_e^2 M_A^2 }
\right),
\nn\\
A_7 =&
\br{ x_1 x_4 - y_1 y_4 } \br{ p_{12} + p_{23} - w_2 } +
\nn\\&
+ \br{ x_1 x_2 - y_1 y_2 } \br{ p_{14} + p_{34} - w_4 } -
\nn\\&
 - 2 M_A^2 \br{ 2 \br{ x_1 x_3 - y_1 y_3 } + x_1^2 - y_1^2 },
\nn\\
A_8 =&
-\br{ x_3 x_4 - y_3 y_4 } \br{p_{12} + p_{23} + w_2} -
\nn\\&
-\br{ x_2 x_3 - y_2 y_3 } \br{p_{14} + p_{34} + w_4} +
\nn\\&
+ 2 M_A^2 \br{ 2 \br{ x_1 x_3 - y_1 y_3 } + x_3^2 - y_3^2 },
\nn\\
A_9 = &
\br{ x_1^2 - y_1^2 }
\left(
	p_{12} p_{34} + p_{23} p_{14} -
\right.
\nn\\&
- w_2 p_{34} + 2 w_3 M_A^2 - w_4 p_{23} -
\nn\\&
\left.
- 2 \br{ M_A^2 p_{13} + m_e^2 p_{24} - 4 m_e^2 M_A^2 }
\right),
\nn\\
A_{10} =&
\br{ x_3^2 - y_3^2 }
\left(
	p_{12} p_{34} + p_{23} p_{14} -
\right.
\nn\\&
- 2 w_1 M_A^2 + w_2 p_{14} + w_4 p_{12} -
\nn\\&
\left.
- 2 \br{ M_A^2 p_{13} + m_e^2 p_{24} - 4 m_e^2 M_A^2 }
\right),
\nn\\
A_{11} =&
2 ( x_2 x_4 - y_2 y_4 ).
\nn\\
A_{12} =&
\br{ x_1 y_3 + x_3 y_1 }
\left(
2 w_2 w_4
- 2 \br{ p_{14} p_{23} + p_{12} p_{34} }
\right. -
\nn\\&
- w_2 \br{ p_{14} - p_{34} }
- w_4 \br{ p_{12} - p_{23} } +
\nn\\&
+ \left.
4 M_A^2 \br{ p_{13} - 4 m_e^2 } +
4 m_e^2 p_{24}
\right),
\nn\\
A_{13} =&
\br{ x_1 y_4 + x_4 y_1 }\br{ p_{12} + p_{23} - w_2 } +
\nn\\&
+ \br{ x_1 y_2 + x_2 y_1 } \br{ p_{14} + p_{34} - w_4 } -
\nn\\&
- 4 M_A^2 \br{ x_1 y_1 + x_1 y_3 + x_3 y_1 },
\nn\\
A_{14} =&
- \br{ x_3 y_4 + x_4 y_3 }\br{ p_{12} + p_{23} + w_2 } -
\nn\\&
- \br{ x_2 y_3 + x_3 y_2 } \br{ p_{14} + p_{34} + w_4 } +
\nn\\&
+ 4 M_A^2 \br{ x_1 y_3 + x_3 y_1 + x_3 y_3 },
\nn\\
A_{15} =&
2 x_1 y_1 \left(
p_{12} p_{34} + p_{14} p_{23} - p_{34} w_2 - p_{23} w_4
\right.-
\nn\\&
-
\left. 2 M_A^2 \br{ p_{13} - w_3 - 4 m_e^2 } - 2 m_e^2 p_{24}
\right),
\nn\\
A_{16} =&
2 x_3 y_3 \left(
p_{12} p_{34} + p_{14} p_{23} + p_{14} w_2 + p_{12} w_4
\right.-
\nn\\&
-
\left. 2 M_A^2 \br{ p_{13} + w_1 - 4 m_e^2 } - 2 m_e^2 p_{24}
\right),
\nn\\
A_{17} =&
2 ( x_2 y_4 + x_4 y_2 ).
}
Here we used additional abbreviations for shortening:
$x_i \equiv 2\br{\varepsilon_1 p_i}$ and $y_i \equiv 2\br{ \varepsilon_2 p_i}$.

\subsection{Circular polarization}

In this section we present the quantity ${\cal S}_{P_3}$ from (\ref{eq.def.calSP3})
for the cross sections of emission of the circularly polarized photon:
\eq{
&
{\cal S}_{P_3} =
\frac{A_{18}}{w_1 w_3}
+
2\br{\frac{A_{19}}{w_1}
+
2\frac{A_{20}}{w_3}}
+
4\br{\frac{A_{21}}{w_1^2}
+
\frac{A_{22}}{w_3^2}},
}
where
\eq{
A_{18} =&
\hspace{4.3mm}4 E_{2kxy} (2 a_{1k} p_{23} - m_e^2(a_{12} + 3 a_{14}) ) -
\nn\\&
- 4 E_{13xy} ((a_{13} + a_{14}) p_{23} + a_{12} p_{34} +
\nn\\&\hspace{30mm}+ (a_{13} - a_{1k}) w_2 + a_{12} m_e^2) +
\nn\\&
+ 4 E_{1kxy} (2 a_{13} M_A^2 + 2 a_{1k} p_{23} - a_{13} (p_{23} + p_{24}) -
\nn\\&\hspace{5mm}- (a_{13} + a_{14} - a_{1k}) w_2 - a_{12} w_4 - 2 a_{12} m_e^2) +
\nn\\&
+ 2 E_{3kxy} (a_{14} p_{12} + a_{12} p_{14} - 4 a_{1k} p_{23} +
\nn\\&\hspace{30mm}+ 4 a_{12} w_4 + 2 a_{12} m_e^2) +
\nn\\&
+ 4 E_{12xy} ((a_{12} + a_{14}) m_e^2 + a_{13} (p_{23} + p_{34}) +
\nn\\&\hspace{20mm}+ (a_{13} - a_{1k}) (w_2 + w_4)) +
\nn\\&
+ 2 E_{3xya} (-2 ((p_{13} + p_{14}) p_{23} + p_{12} p_{34}) -
\nn\\&\hspace{5mm}- (p_{13} + p_{14} + 2 m_e^2) w_2 - p_{12} w_4 - 2 m_e^2 p_{12}) +
\nn\\&
+ 2 E_{1xya} (2 p_{13} p_{23} + (p_{13} - p_{34} + 2 m_e^2) w_2 -
\nn\\&\hspace{15mm}- p_{23} w_4 + 2 m_e^2 (p_{12} - 2 p_{24})) -
\nn\\&
- 2 E_{kxya} (2 p_{13} p_{23}+ (p_{13} + 2 (p_{14} +  m_e^2)) w_2 +
\nn\\&\hspace{15mm}+ 2 (p_{12} - 2 p_{23}) w_4 + 2 m_e^2 (p_{12} + 4 p_{24})) +
\nn\\&
+ 2 E_{2xya} (2 p_{13} (p_{23} + p_{34}) + (p_{13} + 2 m_e^2) w_2 +
\nn\\&\hspace{15mm}+ (p_{13}+ 6 m_e^2) w_4 + 2 m_e^2 (p_{12} + p_{14})) -
\nn\\&
- E_{23ya} (4 (p_{23} + p_{34}) x_1 + 2 (p_{12} + p_{14}) x_3 -
\nn\\&\hspace{15mm}- w_2 (x_1 - 2 x_3) - w_4 (x_1 - 6 x_3)) -
\nn\\&
- 2E_{12ya} ((p_{23} + p_{34}) x_3 - w_2 (2 x_1 + x_3) -
\nn\\&\hspace{15mm}- w_4 (2 x_1 + x_3) - 2 m_e^2 (x_2 + x_4)) -
\nn\\&
- E_{13ya} (4 p_{23} x_1 + 2 (p_{12} - p_{23} - 2 p_{24}) x_3 +
\nn\\&\hspace{10mm}+ w_2 (3 x_1 + 4 x_3 - x_4) - w_4 x_2 + 4 m_e^2 x_2) -
\nn\\&
- E_{2kya} ((4 (p_{12} + p_{14}) - p_{23} - p_{34}) x_1 +
\nn\\&\hspace{10mm}+ 2 p_{13} (x_2 + x_4) + (p_{12} + p_{14} + 8 p_{23}) x_3 +
\nn\\&\hspace{10mm}+ 4 (w_2 + w_4) x_3 + 4 m_e^2 (x_2 + 3 x_4)) -
\nn\\&
- E_{1kya} (32 M_A^2 x_3 + (2 p_{13} + p_{34}) x_2 - 10 p_{24} x_3 +
\nn\\&\hspace{5mm}+ p_{12} (4 x_1 + x_3) - p_{23} (x_1 - 6 x_3 - x_4) +
\nn\\&\hspace{5mm}+ 2 w_2 (2 x_1 + 3 x_3 - 2 x_4) - 4 w_4 x_2 + 8 m_e^2 x_2) -
\nn\\&
- E_{3kya} (16 M_A^2 (x_1 - x_3) + 5 p_{23} x_1 -
\nn\\&\hspace{10mm}- 2 (p_{13} + p_{14}) x_2 - 8 (p_{23} - p_{24}) x_3 -
\nn\\&\hspace{10mm}- p_{12} (4 x_1 - x_3 + 2 x_4) -
\nn\\&\hspace{10mm}- w_2 (x_1 + 2 x_3 + 4 x_4) - 4 w_4 x_2 - 4 m_e^2 x_2) -
\nn\\&
- 2 E_{123y} ((a_{12} + a_{14}) x_3 + 2 (p_{23} + p_{34}) x_a) -
\nn\\&
- 2 E_{12ky} (2 a_{1k} x_3 + a_{12} (2 x_1 - x_3) +
\nn\\&\hspace{10mm}+ a_{14} (2 x_1 - x_3) + a_{13} (2 (x_2 + x_4) - x_3) -
\nn\\&\hspace{10mm}- (p_{23} + p_{34}) x_a) +
\nn\\&
+ E_{13ky} (8 M_A^2 x_a + (4 (a_{13} + a_{1k}) - a_{14}) x_2 -
\nn\\&\hspace{10mm}- a_{12} (x_1 + 6 x_3 + x_4) -
\nn\\&\hspace{10mm}- (p_{12} + 6 p_{23} + 2 p_{24} - 4 w_2) x_a) +
\nn\\&
+ E_{23ky} (4 a_{1k} x_2 + 2 a_{13} x_3 - a_{12} (5 x_1 + 2 x_3) -
\nn\\&\hspace{10mm}- a_{14} (5 x_1 + 6 x_3) - 4 a_{1k} (x_3 + x_4) -
\nn\\&\hspace{10mm}-  (p_{12} + p_{14} - 4 (w_2 + w_4)) x_a) +
\nn\\&
+ E_{13ka} (4 ( x_1 y_2 - x_2 y_1) - 7 ( x_2 y_3 - x_3 y_2)) +
\nn\\&
+ E_{23xa} (4 (p_{23} + p_{34}) y_1 + 2 (p_{12} + p_{14}) y_3 -
\nn\\&\hspace{10mm}- w_2 (y_1 - 2 y_3) - w_4 (y_1 - 6 y_3)) +
\nn\\&
+ E_{13xa} (4 p_{23} y_1 + 2 (p_{12} - p_{23} - 2 p_{24}) y_3 +
\nn\\&\hspace{10mm}+ w_2 (3 y_1 + 4 y_3 - y_4) - w_4 y_2 + 4 m_e^2 y_2) +
\nn\\&
+ 2 E_{12xa} ((p_{23} + p_{34}) y_3 - w_2 (2 y_1 + y_3) -
\nn\\&\hspace{10mm}- w_4 (2 y_1 + y_3) - 2 m_e^2 (y_2 + y_4)) -
\nn\\&
- 2( E_{123a} + 2 E_{23ka} ) ((x_2 + x_4) y_3 - x_3 (y_2 + y_4)) +
\nn\\&
+ E_{12ka} ((x_2 + x_4) (4 y_1 + y_3) -
\nn\\&\hspace{30mm}- (4 x_1 + x_3) (y_2 + y_4)) +
\nn\\&
+ E_{1kxa} (2 p_{13} y_2 + p_{34} y_2 + p_{23} (-y_1 + 6 y_3 + y_4) -
\nn\\&\hspace{7mm}- 10 p_{24} y_3 + p_{12} (4 y_1 + y_3) + 32 M_A^2 y_3 +
\nn\\&\hspace{7mm}+ 2 w_2 (2(y_1 - y_4) + 3 y_3) - 4 w_4 y_2 + 8 m_e^2 y_2) +
\nn\\&
+ E_{2kxa} ((4 (p_{12} + p_{14}) - p_{23} - p_{34}) y_1 +
\nn\\&\hspace{10mm}+ (p_{12} + p_{14} + 8 p_{23} + 4 (w_2 + w_4)) y_3 +
\nn\\&\hspace{10mm}+ 2 p_{13} (y_2 + y_4) + 4 m_e^2 (y_2 + 3 y_4)) +
\nn\\&
+ E_{3kxa} (16 M_A^2 (y_1 - y_3) + 5 p_{23} y_1 -
\nn\\&\hspace{9mm}- 2 (p_{13} + p_{14}) y_2 - 8 (p_{23} - p_{24}) y_3 -
\nn\\&\hspace{9mm}- p_{12} (2 (2 y_1 + y_4) - y_3) -
\nn\\&\hspace{9mm}- w_2 (y_1 + 2 (y_3 + 2 y_4))- 4 w_4 y_2 - 4 m_e^2 y_2) +
\nn\\&
+ 2 E_{12kx} (2((a_{12} + a_{14}) y_1 + a_{13} y_2 + a_{13} y_4 ) -
\nn\\&\hspace{10mm}- (a_{12} + a_{13} + a_{14} - 2 a_{1k}) y_3 -
\nn\\&\hspace{10mm}- (p_{23} + p_{34}) y_a) +
\nn\\&
+ 2 E_{123x} ((a_{12} + a_{14}) y_3 + 2 (p_{23} + p_{34}) y_a) +
\nn\\&
+ E_{13kx} ((a_{14} - 4 (a_{13} + a_{1k})) y_2 +
\nn\\&\hspace{10mm}+ a_{12} (y_1 + 6 y_3 + y_4) - 8 M_A^2 y_a +
\nn\\&\hspace{10mm}+ (p_{12} + 2( 3 p_{23} + p_{24} - 2 w_2)) y_a) -
\nn\\&
- E_{23kx} (2 a_{13} y_3 - a_{12} (5 y_1 + 2 y_3) -
\nn\\&\hspace{10mm}- a_{14} (5 y_1 + 6 y_3) + 4 a_{1k} (y_2 - y_3 - y_4) -
\nn\\&\hspace{10mm}- (p_{12} + p_{14} - 4 (w_2 + w_4)) y_a) +
\nn\\&
+ E_{123k} (x_a (y_2 + 2 y_3 + y_4) - (x_2 + 2 x_3 + x_4) y_a),
\nn
\\
A_{19} =&
-4 a_{12} (E_{1kxy} + E_{2kxy} - E_{3kxy}) -
\nn\\&
- E_{kxya} (p_{12} + 2 p_{23}) - E_{12ya} (x_2 - x_3 + x_4) +
\nn\\&
+ (E_{1xya} - E_{3xya}) (p_{12} + 2 (p_{23} + 2 M_A^2)) +
\nn\\&
+ E_{2xya} (p_{12} + p_{14} + 2 (p_{23} + p_{34})) +
\nn\\&+ ( E_{13ya} - E_{1kya} + 2 E_{3kya}) x_2 + E_{23ya} x_3 -
\nn\\&- E_{2kya} (2(x_2 - x_4) - x_3) +
\nn\\&+ 2 ( E_{12ky} + E_{23ky} ) x_a - E_{23xa} y_3 -
\nn\\&- (E_{13xa} - E_{1kxa} + 2 E_{3kxa}) y_2 +
\nn\\&+ E_{12xa} (y_2 - y_3 + y_4) +
\nn\\&+ E_{2kxa} (2 (y_2 - y_4) - y_3) - 2 (E_{12kx} + E_{23kx}) y_a,
\nn
\\
A_{20} =&
- (E_{1xya} - E_{kxya}+ E_{2xya} - E_{3xya}) w_2 -
\nn\\&\hspace{25mm}- w_4 E_{2xya} + 2 M_A^2E_{3xya},
\nn
\\
A_{21} =&
-2 ((a_{12} + a_{14}) E_{2kxy} - a_{12} E_{3kxy} ) m_e^2 +
\nn\\&
+ E_{13xy} (a_{1k}(p_{23} - 2 M_A^2) - a_{12} m_e^2) +
\nn\\&
+ E_{1kxy} (a_{12} p_{34} + (a_{14} + a_{1k}) p_{23} -
\nn\\&\hspace{20mm}-2 a_{13} M_A^2 - 3 a_{12} m_e^2) -
\nn\\&
- E_{12xy} (a_{1k} (p_{23} + p_{34}) - (a_{12} + a_{14}) m_e^2) +
\nn\\&
+ E_{3xya} m_e^2 (4 M_A^2 - p_{12} + 2 ( 2 p_{23} - w_2 )) +
\nn\\&
+ E_{1xya} (2 (p_{12} p_{34} + p_{14} p_{23}) - 2 M_A^2 (p_{13} - 2 w_3) +
\nn\\&\hspace{20mm}
+ m_e^2 ( p_{12} - 2 (2 p_{23} + p_{24} -  w_2))) +
\nn\\&
+ E_{2xya} m_e^2 (p_{12} + p_{14} -
\nn\\&\hspace{20mm}-2 (2 (p_{23} + p_{34}) - w_2 - w_4)) -
\nn\\&
- E_{kxya} ((p_{14} - 2 w_4) p_{23} + (p_{12} - 2 w_2) p_{34} +
\nn\\&\hspace{9mm} + 2 M_A^2 (p_{13} + 2 w_3) +
\nn\\&\hspace{9mm} + m_e^2 (p_{12} - 2 ( 2(p_{23} - p_{24}) - w_2 - 4 M_A^2))) +
\nn\\&
+ E_{13ya} (2 x_1 (p_{23} + M_A^2) - m_e^2 x_2) -
\nn\\&
- E_{3kya} ( x_1 (p_{23} +2 M_A^2) - 2 m_e^2 x_2) +
\nn\\&
+ E_{1kya} ((3 p_{23} - 2 p_{24}) x_1 + 4 M_A^2 (x_1 - x_3) -
\nn\\&\hspace{58mm}- 3 m_e^2 x_2) +
\nn\\&
+ E_{2kya} (x_1 (p_{23} + p_{34}) - 2 m_e^2 (x_2 + x_4)) -
\nn\\&
- E_{12ya} (2 x_1 (p_{23} + p_{34}) - m_e^2 (x_2 + x_4)) +
\nn\\&
+ E_{12ky} (p_{23} + p_{34}) x_a
-  E_{13ky} (p_{23} - 2 M_A^2) x_a +
\nn\\&
+ E_{3kxa} ((p_{23} + 2 M_A^2) y_1 - 2 m_e^2 y_2) -
\nn\\&
- E_{13xa} (2 (p_{23} + M_A^2) y_1 - m_e^2 y_2) -
\nn\\&
- E_{1kxa} ((3 p_{23} - 2 p_{24}) y_1 + 4 M_A^2 (y_1 - y_3) -
\nn\\&\hspace{58mm}- 3 m_e^2 y_2) +
\nn\\&
+ E_{12xa} (2 (p_{23} + p_{34}) y_1 - m_e^2 (y_2 + y_4)) -
\nn\\&
- E_{2kxa} ((p_{23} + p_{34}) y_1 - 2 m_e^2 (y_2 + y_4)) -
\nn\\&
- E_{12kx} (p_{23} + p_{34}) y_a
+ E_{13kx} (p_{23} - 2 M_A^2) y_a,
\nn
\\
A_{22} =&
E_{3kxy} (2 (a_{14} w_2 + a_{12} w_4) - 4 a_{1k} M_A^2) +
\nn\\&
+ (E_{12ky} + 2E_{23ky})(a_{12} + a_{14}) x_3 -
\nn\\&
- (E_{12kx} + 2 E_{23kx})(a_{12} + a_{14}) y_3 +
\nn\\&
+ a_{12} (E_{13ky} x_3 - E_{13kx} y_3) +
\nn\\&
+ (E_{2kya} x_3 - E_{2kxa} y_3) (p_{12} + p_{14}) +
\nn\\&
+ (E_{13ya} x_3 - E_{13xa} y_3) w_2 +
\nn\\&
+ (E_{3kya} x_3 - E_{3kxa} y_3) (4 M_A^2 - p_{12}) +
\nn\\&
+ (E_{1kya} x_3 - E_{1kxa} y_3) (p_{12} + w_2) -
\nn\\&
- (E_{12ya} x_3 - E_{12xa} y_3) (w_2 + w_4) +
\nn\\&
+ E_{13ka} (x_2 y_3 - x_3 y_2) -
\nn\\&
- E_{12ka} ((x_2 + x_4) y_3 - x_3 (y_2 + y_4)),
}
where we used the following invariants:
\eq{
x_a &\equiv 2\br{e_1 a_1}, &\qquad y_a &= 2\br{e_2 a_1},	\nn\\
a_{12} &\equiv 2\br{a_1 p_2},&\qquad
a_{13} &\equiv 2\br{a_1 p_3},\\
a_{14} &\equiv 2\br{a_1 p_4},&\qquad
a_{1k} &\equiv 2\br{a_1 k}.\nn
}
The latter expressions for the traces contain many
contractions of vectors with the antisymmetric Levi-Civita tensor in the form:
\eq{
	E_{a b c d}
	\equiv
	m_e \, \varepsilon_{\mu \nu \alpha \beta}
	a^\mu b^\nu c^\alpha d^\beta.
	\label{eq.EpsNotation}
}
These contractions can be written out using the explicit components of the 4-vectors from Table~\ref{table.VectorComponents}.

Below we give an example of calculation of the quantity $E_{a b c d}$ in order to show why we inserted the
mass of electron $m_e$ into that notation.
The formal proportionality of the total expression for ${\cal S}_{P_3}$ to the mass of electron $m_e$
does not mean that the whole contribution is small. In fact, in the case of
the longitudinally polarized initial electron beam if one of the vectors in the quantity $E_{a b c d}$ is
the vector $a_1$, then this contribution is not suppressed by the electron mass.
So, in case of the quantity $E_{2xya}$ we have
\eq{
E_{2xya}
&=
m_e
\brm{
\begin{array}{cccc}
{p_2}_0 & {p_2}_x & {p_2}_y & {p_2}_z \\
{\varepsilon_1}_0 & {\varepsilon_1}_x & {\varepsilon_1}_y & {\varepsilon_1}_z \\
{\varepsilon_2}_0 & {\varepsilon_2}_x & {\varepsilon_2}_y & {\varepsilon_2}_z \\
{a_1}_0 & {a_1}_x & {a_1}_y & {a_1}_z \\
\end{array}
}
=
\nn\\
&=
\brm{
\begin{array}{cccc}
M_A & 0 & 0 & 0 \\
0 & C_{1k} \cos\varphi_k & C_{1k} \sin\varphi_k & -S_{1k} \\
0 & -\sin\varphi_k & \cos\varphi_k & 0 \\
\lambda_1 \brm{\vv{p_1}} & 0 & 0 & \lambda_1 E_1 \\
\end{array}
}.
\nn
}
And evaluation of this determinant gives
\eq{
	E_{2xya}
	=
	\lambda_1  C_{1k} M_A E_1.
	\label{eq.EpsContraction}
}

\vspace{8mm}


\end{document}